\documentclass[aps,prb,twocolumn,showpacs,citeautoscript,superscriptaddress]{revtex4-1}

\usepackage{amsmath,amssymb}
\usepackage{bm}
\usepackage{graphicx}
\usepackage{epstopdf} 
\usepackage{latexsym} 
\usepackage{subfigure}
\usepackage{color}
\usepackage{dsfont} 
\usepackage{wasysym} 

\newcommand{\mc}{\cal}

\begin{document}

\title{Infinite critical boson non-Fermi liquid on heterostructure interfaces}
\author{Xiao-Tian Zhang}
\affiliation{Kavli Institute for Theoretical Sciences, University of Chinese Academy of Sciences, Beijing, China}
\affiliation{Department of Physics and HKU-UCAS Joint Institute for Theoretical and Computational Physics at Hong Kong, 
The University of Hong Kong, Hong Kong, China}
\author{Gang Chen}
\email{gangchen@hku.hk}
\affiliation{International Center for Quantum Materials, School of Physics, Peking University, Beijing 100871, China}
\affiliation{Department of Physics and HKU-UCAS Joint Institute for Theoretical and Computational Physics at Hong Kong, 
The University of Hong Kong, Hong Kong, China}
\affiliation{The University of Hong Kong Shenzhen Institute of Research and Innovation, Shenzhen 518057, China}
\affiliation{Collaborative Innovation Center of Quantum Matter, Beijing 100871, China}

\begin{abstract}
We study the emergence of non-Fermi liquid on heterostructure interfaces where there exists an infinite number of critical boson modes accounting for the magnetic fluctuations in two spatial dimensions. The interfacial Dzyaloshinskii-Moriya interaction naturally arises in magnetic interactions due to the absence of inversion symmetry, resulting in a degenerate contour for the low-energy bosonic modes in the momentum space which simultaneously becomes critical near the magnetic phase transition. The itinerant electrons are scattered by the critical boson contour via the Yukawa coupling. When the boson contour is much smaller than the Fermi surface, it is shown that, there exists a regime with a dynamic critical exponent ${z=3}$ while the boson contour still controls the low-energy magnetic fluctuations. Using a self-consistent renormalization calculation for this regime, we uncover a prominent non-Fermi liquid behavior in the resistivity with a characteristic temperature scaling power. These findings open up new avenues for understanding boson-fermion interactions and the novel fermionic quantum criticality. 
\end{abstract}

\maketitle

\section{Introduction}

Heterostructures at various interfaces have been fascinating hunting 
grounds for emergent phenomena that are absent in the bulk 
constituents~\cite{Heber2009,Mannhart2010,Boris2011}. One of
the most well-known emergent phenomena that were discovered 
on the heterostructure interfaces were the quantum Hall effects 
on the GaAs/AlGaAs semiconductor heterostructure~\cite{PhysRevLett.48.1559}
and the coexisting superconductivity and ferromagnetism (FM)
on the LaAlO$_3$/SrTiO$_3$ oxide interface~\cite{Kalisky_2012}. 
More recently, tremendous efforts have been devoted into 
investigation on the oxide interface formed by two structurally 
and chemically distinct transition-metal compounds~\cite{Hwang2012,Chakhalian2012}.
The interface is associated with the breaking of the spatial inversion 
symmetry by design. The lower dimensionality reduces the 
bandwidth and leads to an enhancement of electron correlations,
thus, giving rise to interesting many-body phenomena 
including magnetism, metal-insulator transition, 
unconventional superconductivity~\cite{CRMP2014}.
From the experimental point of view, the heterostructure geometry
enables otherwise unattainable manipulation with the external fields
and probing measurements. For instance, 
the detection of the atomic-scale magnetic skyrmion 
at Fe ultra-thin film~\cite{Zhang2015} on the Ir(111) surface 
by scanning tunnelling microscopy~\cite{Heinze2011}
and the electric-field-driven switching of individual magnetic 
skyrmions~\cite{Hsu2016}. Recent technical advances 
in the atomic-scale fabrication of oxide heterostructures
and experimental probe of electronic and magnetic orders 
in ultra-thin films~\cite{Heinze2011,Fert2013,Hsu2016,King2014,Liu2013} 
have rendered the interfaces a promising platform 
for the study of various electron correlation phenomena.

One prominent example is the heterostructure interface of rare-earth 
nickelate oxides~\cite{Boris2011,King2014} that is reported to be a 
quantum critical metal owing to its proximity to an antiferromagnetic (AFM) 
quantum critical point (QCP)~\cite{Liu2020,Liu2013,Guo2018}.
The critical fluctuation of the magnetic order parameter around the QCP,
{\sl i.e.} the antiferromagnetic-type fermion criticality~\cite{Abanov2004,Abrahams2012,SSLee2015,SSLee2017}, 
is known as the critical boson.
The electrons near the Fermi surface, when coupled to such critical bosons,
are strongly scattered and enter the respective non-Fermi liquid  
phases or regimes~\cite{Hertz1976,Millis1993,Moriya1973,Moriya1973B}.
The non-Fermi liquid behaviors are observed by the transport measurements 
at low temperatures~\cite{Jaramillo2014,Mikheev2015,Liu2020}
that do not exist in the bulk counterparts~\cite{Liu2013}.
The resistivity shows a peculiar temperature dependence 
${\Delta \rho(T) \sim T^{\alpha}}$, and provides an evidence 
for the emergence of unusual metallic non-Fermi liquid 
phases with ${\alpha \ne 2}$. 
More exotically, the scaling exponent is tunable within the interval ${1\le \alpha \le2}$
upon applying pressure by strain~\cite{Zhou2005,Mikheev2015,Liu2020}.
The burst of the experimental evidence on correlated phenomena in oxide 
heterostructures urge for theoretical frameworks and understanding 
of the underlying mechanism.


\begin{figure}[b] 
	\centering
	\includegraphics[width=8.5cm]{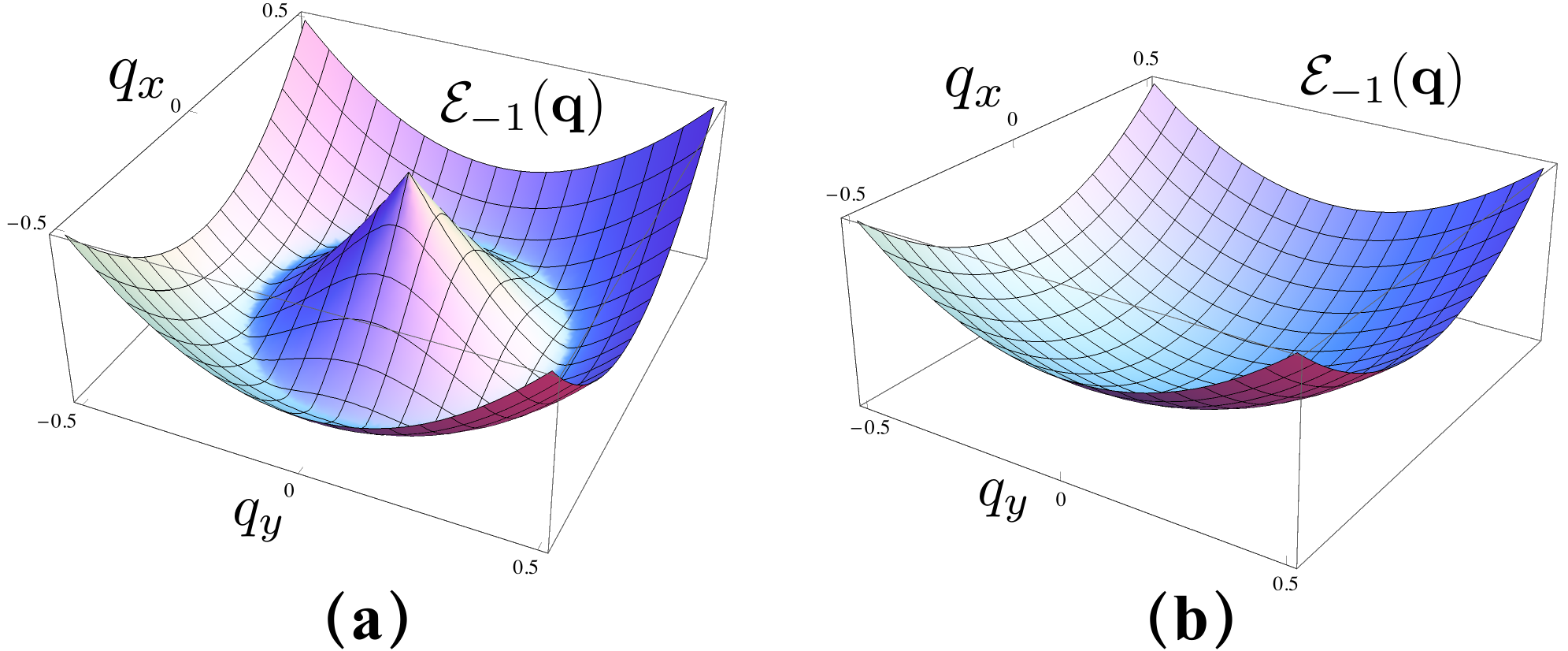}
	\caption{(Color online.) The comparison of the representative 
	low-energy boson dispersions
	with and without the DM interaction, where
	the left contains the degenerate contour minima. 
	}
	\label{fig1}
\end{figure}

We consider the heterostructure interfaces with both
itinerant electrons and magnetism. 
The itinerant electrons can arise from the interfacial 
reconstruction and the charge transfer from the bulk
for polar-non-polar interfaces~\cite{Hwang2012},
or simply from the metallic side of the heterostructure. 
The magnetism comes from the Mott insulating inside of the heterostructure
or arise from the enhanced correlation at the interfaces. 
One of the key ingredients at many interfaces, as mentioned,
is the breaking of the spatial inversion symmetry. 
While this renders a Rashba spin-orbit coupling
for the itinerant electrons, the Fermi surface is most likely 
still present and the itinerant electrons remain in 
the metallic phase. For magnetism, however, 
one immediate effect is the introduction of the 
Dzyaloshinskii-Moriya(DM) interaction between 
the magnetic degrees of freedom~\cite{RMP2017,Dzyaloshinsky1958,Moriya1960}.
With the DM interaction, the interfacial magnetism can display 
a distinct behavior around the quantum phase transition.
In contrast to the conventional AFM quantum criticality,
the magnetic order parameter fluctuation is dominated by 
the critical boson modes on a continuously degenerated 
minima~\cite{Sur2019}. This continuously degenerated bosonic 
minima induced quantum critical phenomena
have aroused extensive attention from the theoretical aspects~\cite{Zhang2023,Lake2021,Tsvelik2021,Ku2021}. 
In this paper, we consider a quantum critical metal with itinerant magnetism
due to strong electron correlation at the two-dimensional (2D) 
oxide interface. The itinerant electrons
are strongly coupled with the critical bosonic contour in 2D.
It is shown that, when the radius boson contour is much smaller than the Fermi momentum,
there exists an interesting crossover regime where the 
dynamical critical exponent ${z=3}$ instead of ${z=2}$. 
Due to the large effective dimensions ${d+z}$ in this crossover regime, 
we adopt a self-consistent renormalization method to tame
the critical fluctuation of the magnetic moments and study 
the physical property of the electrons. We show that the 
electronic state enters an unconventional non-Fermi liquid by evaluating 
the temperature dependence of the resistivity at low energy. 
The interfacial magnetism can be deciphered by the neutron 
scattering experiment whose spectral weight is proposed to 
be divergent around the critical boson contour. Finally, we 
analyze the effect of the external magnetic field and the property 
of the field-tuned QCP.

The structure of paper is organized as follows.
In Sec.~\ref{sec2}, we introduce the field-theoretical model for the coupled  bosons and fermions, 
and demonstrate the presence of the critical boson contour 
and its impact on the low-energy dynamics of bosons.
In Sec.~\ref{sec3}, we consider the fluctuations in the vicinity of the
critical bosonic contour by means of scaling analysis at tree level 
and self-consistent renormalization at one-loop level.
In Sec.~\ref{sec4}, we analyze the non-Fermi liquid behavior in transport
by evaluating the quasiparticle lifetime.
Finally, in Sec.~\ref{sec5}, we discuss on the effect of external 
magnetic field and the case where electronic dispersion shrinks to discrete band touchings.

\begin{figure}[htbp] 
	\centering
	\includegraphics[width=8cm]{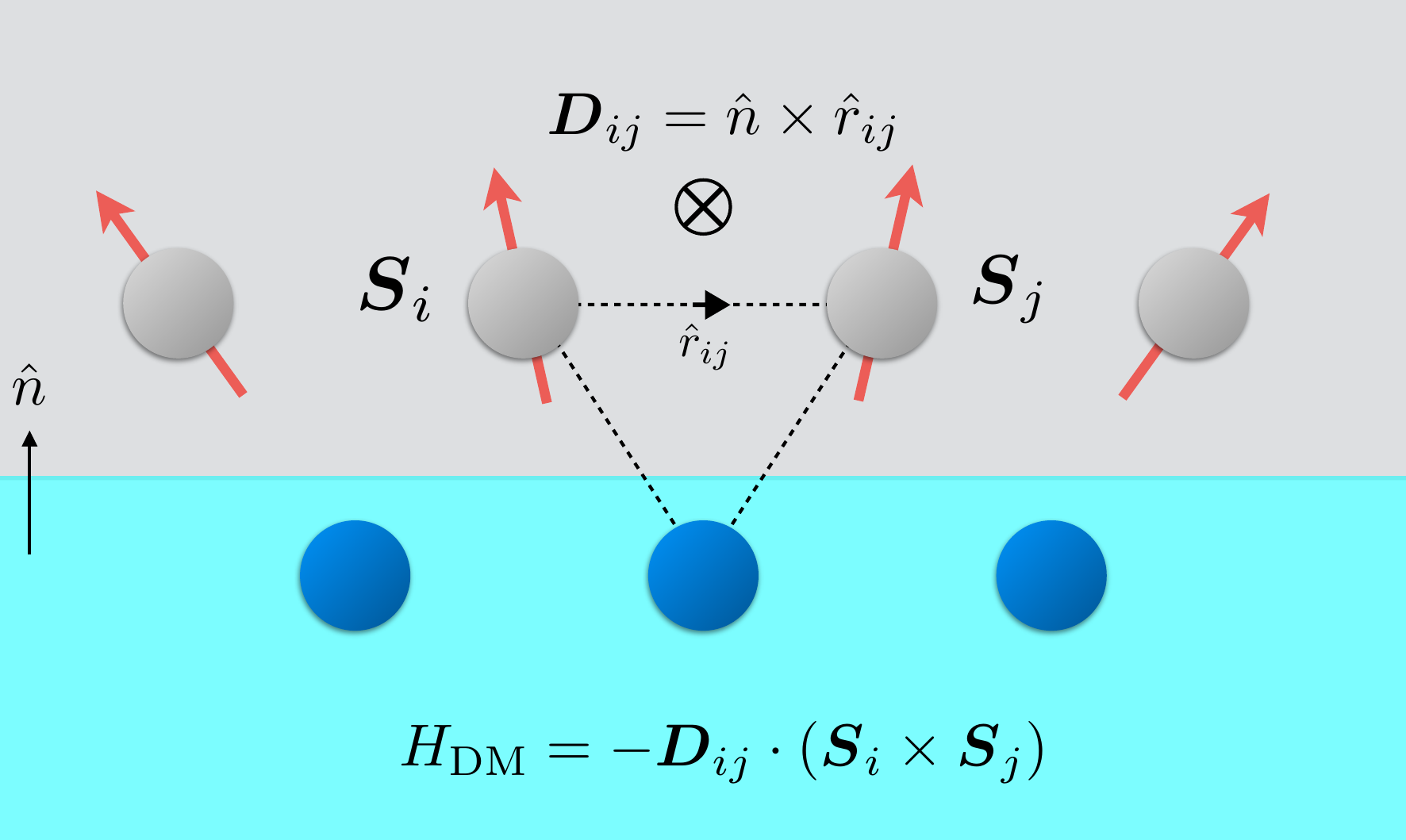}
	\caption{(Color online.) Schematic illustration for generating mechanism
	of the interfacial DM interaction. The magnetic moments in the upper(grey) layer
	is coupled to electrons from the lower(cyan) one.
	Interfaces between magnetic materials and materials with large spin-orbit interactions 
	gives rise to interfacial DM interactions. 
	The interfacial DM interaction Hamiltonian takes a form 
	$H_{\rm DM} = -{\bm D}_{ij}\cdot ({\bm S}_i \times {\bf S}_j)$.
	${\bm D}_{ij}$ is the DM vector originating from the RKKY process
	involving the electrons with large SOC from the lower layer.
	$\hat{n}$ and $\hat{r}_{ij}$ are the directional vectors along the
	normal direction of the interface 
	and connecting adjacent spins ${\bm S}_i, {\bm S}_j$, respectively.
	By fixing the interfacial norm as ${\hat n}={\hat z}$, 
	${\bm D}_{ij} =( \hat{n}\times {\hat r}_{ij} )D$ lies within the interfacial plane.
	The DM interaction is formulated as $\sim (\hat{z} \times {\hat x}) \cdot ({\bf S}\times \partial_x {\bf S}) + (\hat{z} \times {\hat y}) \cdot ({\bf S}\times \partial_y {\bf S}) $.
	}
	\label{DMI}
\end{figure}

\section{Critical boson contour}
\label{sec2}

The interfacial system that we are considering includes 
both itinerant electrons and magnetic degrees of freedom.
Thus, the model for the system naturally contains three parts.
The first part describes the kinetic energy of the itinerant electrons,
the second part describes the Kondo-like coupling between the 
itinerant electron and the magnetic moment, and the third part 
describes the magnetic moments. Since we assume a Fermi liquid metal 
for the itinerant electron sector before incorporating the coupling with the magnetic sector 
and the electron interactions would merely renormalize the kinetic part,
then we can safely start with a renormalized electron kinetic energy. 
 This fermion-boson coupled model takes the following form,
\begin{eqnarray}
 {\cal L}[f^\dagger,f;\phi] & = & 
  \sum_{i,\alpha} f^\dagger_{i\alpha}(\partial_\tau -\mu) f_{i\alpha}^{} 
- \sum_{ij,\alpha} t_{ij}^{} f^\dagger_{i\alpha} f_{j\alpha}^{}
\nonumber 
\\
& + & g \sum_{i,\alpha\beta} f^\dagger_{i\alpha} \vec{\sigma}_{\alpha\beta}^{} 
               f_{i\beta}^{} \cdot \vec{\phi}_{i}^{}
  + {\cal L}_{\rm B}^{} [\vec{\phi}],
  \label{eq1}
\end{eqnarray}
the first line dictates a tight-binding model for the fermion
hopping on a 2D lattice with $f^\dagger(f)$ the fermion creation 
(annihilation) operator of the itinerant electron. 
The $\vec{\sigma}_{\alpha\beta}$ is the Pauli matrix vector   
with ${\alpha, \beta=1,2}$ being the indices for spin-$1/2$.
The fermionic spin couples to the magnetic order parameter 
$\vec{\phi}_i$ via a Kondo-like Yukawa coupling with a magnitude 
$g$. We adopt a coarse-grained Landau-Ginzburg expansion 
for the magnetic fluctuations,
\begin{equation}
\begin{aligned}
{\cal L}_{\rm B}[\vec{\phi}] = & \int d^2{\bm x} 
\frac{1}{2} 
\Big\{ \vec{\phi} \cdot (r-J\nabla^2) \vec{\phi}
+ \frac{u}{2} (\vec{\phi}^2)^2 \\
&{ + D} \big[  {\hat{y}\cdot (\vec{\phi}\times \partial_x \vec{\phi}) }
-\hat{x}\cdot (\vec{\phi}\times \partial_y \vec{\phi})\big] \Big\}\\
\end{aligned}
\label{phi_4}
\end{equation}
where the first line represents a standard $\phi^4$ theory.
The second is the symmetry-allowed interfacial
DM interaction. Due to the symmetry constraint, 
the DM vector ${\bm D}_{ij}$ is aligned aligned along the normal
${\hat{n}}$-direction of the interface.
In the presence of this DM interaction, the global 
rotational symmetry is broken down to the U(1) rotation symmetry 
with respect to the $\hat{n}$-direction. 
We have assumed a dominant ferromagnetic exchange that  
favors an FM in the absence of DM interaction.
Microscopically, the DM interaction could come 
from either the Ruderman-Kittel-Kasuya-Yosida (RKKY) interaction~\cite{Banerjee_2013} 
via the itinerant electrons or the superexchange interaction 
between the local moments themselves. 
In the former case, which is demonstrated in Fig.~\ref{DMI},
the RKKY interaction could transfer the Rashba spin-orbit coupling of the interfacial
itinerant electrons to the local moments and generates the DM interaction.
By aligning the normal $\hat{n}$-direction along the $\hat{z}$-axis,
the DM interaction on the lattices can be transformed into the continuum 
as dictated in Eq.(\ref{phi_4}).
In the latter case, the superexchange interaction contains the DM interaction
as the consequence of the atomic spin-orbit coupling of the local moments. 
Generically, the DM interaction favors the canting of adjacent moments,
when strong enough, leads to non-collinear and twisted magnetic 
configurations~\cite{Heinze2011,Fert2013}. Experimental evidence 
for the interfacial DM interaction have been accumulating from the observation of magnetic ground state configuration
and corresponding excitation spectrum by means of various spectroscopy~\cite{Heinze2011,Fert2013,Hsu2016,RMP2017}.
The interfacial DM interaction has 
also been verified by the first principal-calculations for 
the Fe/Ir(111) interface~\cite{Dup2014} and so on~\cite{Li2021}.

\begin{figure}[t] 
	\centering
	\includegraphics[width=8cm]{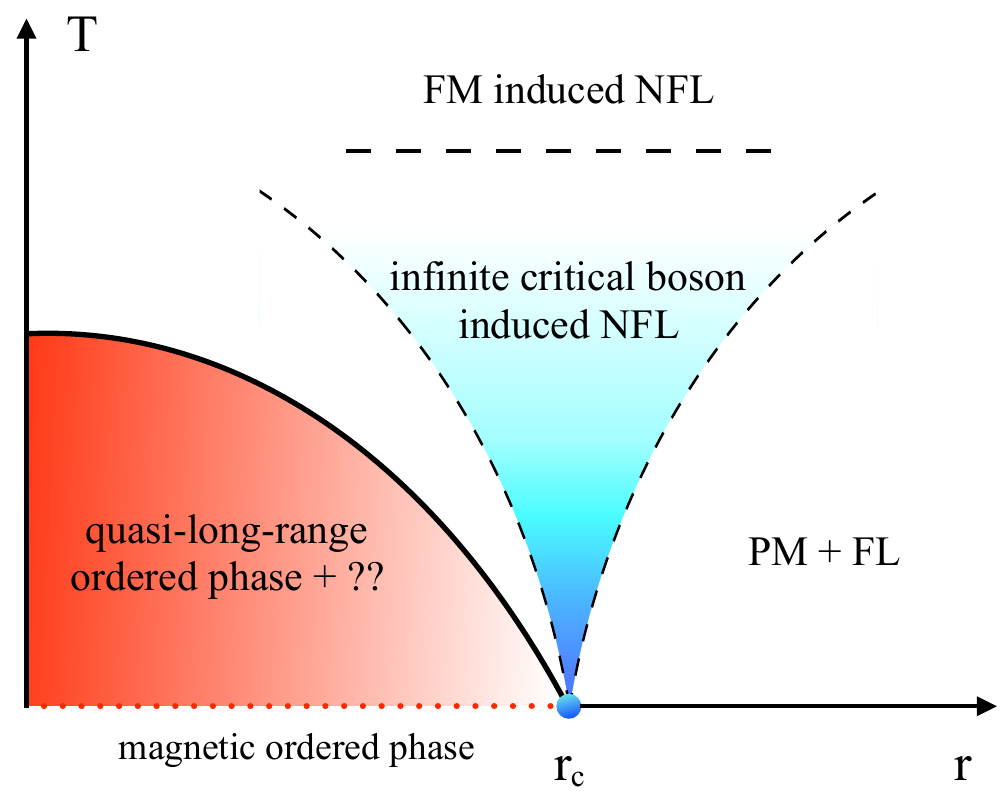}
	\caption{(Color online.) The schematic phase diagram.
	At zero temperature, there exists a QCP at ${r=r_c}$ 
	that separates the zero-temperature magnetic order 
	and the paramagnet (PM). The PM extends to the finite-temperature regime
	where the fermions are in the Fermi liquid (FL) metal phase.
	In contrast, the magnetic order gives way to the quasi-long-range order 
	at finite temperatures that further experiences
	 a Kosterlitz-Thouless-like transition. The QCP mediates 
	 a quantum critical regime at finite temperatures where
	the infinite critical boson induces a novel type of non-Fermi liquid (NFL). 
	The crossover to the conventional FM criticality-induced NFL 
	is experienced at high temperatures.
	The dashed lines mark the crossover from infinite critical boson NFL
	to other quantum critical regimes.		
	}
	\label{pdiagram}
\end{figure}

\subsection{DM interaction and critical boson contour}

The presence of this 2D DM interaction fundamentally modifies 
the physics of the bosonic sector. Without the DM interaction, 
the bosons would condense at the momentum     
${{\bm q} = {\bm 0}}$ with the FM order for ${r=0}$. 
In the presence of the DM interaction, the system reaches a quantum 
critical point (QCP) locating at ${r=r_c}$, and the minima of the boson dispersion 
constitutes a degenerate contour in the 2D reciprocal space with a radius  
$q_0$ (see Fig.~\ref{fig1}). The eigenmodes on the contour corresponding to
non-collinear and twisted magnetic structures.  
To explore the dynamic property of the bosonic sector, 
we follow the established Hertz-Millis scheme~\cite{Hertz1976,Millis1993}
and firstly integrate out the fermionic degree of freedom 
as shown in Fig.~\ref{fig3}(a). This process generates 
a Landau damping term for the bosons that arises 
from the particle-hole excitations around the Fermi level.
The effective boson action takes a form
\begin{eqnarray}
\mc{S}_{\rm B} &= &
 {{{\cal S}_{\rm B}}^{(2)} } 
+ \frac{u}{4}\int d\tau\int d^2{\bm x} \big[\phi^2({\bm x},\tau)\big]^2 ,\\
{{{\cal S}_{\rm B}}^{(2)} } & = & 
\frac{1}{2}\sum_{{\bm q}, i\omega_l} \Pi_{\mu\nu}({\bm q},i\omega_l) 
\phi_{\mu}({\bm q}, i \omega_l) \phi_{\nu}(-{\bm q},-i\omega_l) , 
\label{S_B}
\end{eqnarray}
where ${\mu,\nu=x,y,z}$ label the vector components of 
$\vec{\phi}$, and ${\bm q}$ and ${\omega_l=2\pi l/\beta\ (l\in {\mathbb Z}) }$ 
are the momentum and bosonic Matsubara frequency, respectively.
The renormalized quadratic action, {\sl i.e.} the polarization bubble 
in Fig.~\ref{fig3}(a), takes the following form
\begin{equation}
 \Pi({\bm q},i\omega_l)=  
 \left[\begin{array}{ccc} 
f({\bm q},i\omega_l) & 0 & iDq_x \\
0 & f({\bm q},i\omega_l) & -iDq_y \\
-iDq_x & iDq_y & f({\bm q},i\omega_l) \\
 \end{array}\right] ,
\label{boson_Pi}
\end{equation}
The function on the diagonal entry is given by 
\begin{equation}
f(q,i\omega_l) = r+ J {q}^2 + \delta \Pi(q,i\omega_l).
\end{equation}
where $ \delta \Pi(q,i\omega_l)$ represents the boson self-energy 
correction as derived in Sec.~\ref{sec:Pi}.
Generically, we consider the characteristic form 
$\delta \Pi(q,i\omega_l) =  |\omega_l| /{\it \Gamma}_q$ 
which is known as the Landau damping term. 
We diagonalize the bare part of the quadratic boson action  
and obtain three branches of the eigenmodes,
\begin{eqnarray}
&& {\cal E}_{n}({\bm q},i\omega_l)
= f({\bm q},i\omega_l) +nDq, \quad n=0,\pm 1.
\label{eigen_m}
\end{eqnarray}
The expression of the eigenmodes, particularly the lowest branch 
with ${n=-1}$, determines the nature of the bosonic sector at the 
QCP and the property of the magnetic phase transition.

We propose that the low-energy physics in the 
quantum critical regime is crucially affected even 
by a weak DM interaction (${D\ll J}$). 
As dictated in Eq.~\eqref{S_B} and Eq.~\eqref{boson_Pi}, 
the DM interaction term 
complicates the low-energy theories by 
introducing the vector index into the bosonic sector.
The dispersion of the bosonic modes is modified   
compared to the ${D=0}$ case where the critical 
boson mode reaches its minima only at discrete momenta
such as the FM and AFM criticalities.
In the presence of the interfacial DM interaction, 
there exist infinite critical bosons on continuously degenerate minima.
Let us focus on the lowest eigenmode, {\sl i.e.} the ${n=-1}$ 
branch of Eq.~\eqref{eigen_m}, in the static limit ${\omega_l=0}$.
The dispersion of the lowest ${n=-1}$ branch is now abbreviated as 
${\cal E}_q$ and reads
\begin{equation}
{\cal E}_q= \delta+ J(q-q_0)^2,
\label{E_q}
\end{equation}
which reaches its minimum on a contour 
in the momentum space described by ${q=q_0 \equiv D/(2J)}$. 
The modes on the contour become simultaneously gapless 
when we approach the critical point locating at ${r=r_c= D^2/4J}$.
The representative phase diagram is depicted in Fig.~\ref{pdiagram}. 
On the right corner, the local moment (or bosonic) sector of 
the system realizes a paramagnetic phase with the gapped magnetic 
excitations. On the left corner, the local moment sector develops 
a helical magnetic order that spontaneously breaks the 
U(1) symmetry of the model in Eq.~\eqref{phi_4}. At finite temperatures, 
there is no magnetic order in 2D, and there will be a Kosterlitz-Thouless-like  
transition due to the U(1) symmetry~\cite{Kosterlitz_1973}.
This U(1) symmetry is present due to the simplification of the chosen model.
In reality, it can be broken by other weak anisotropies that will be mentioned in  
Sec.~\ref{sec5}, and then there can be magnetic orders at finite temperatures.
At the quantum phase transition, all the bosonic modes 
on the degenerate contour become critical at the same time, 
and thus we dub the degenerate contour 
``\emph{critical boson contour}'' (CBC).

\subsection{Fermion-boson coupling}
\label{sec:CBC}

\begin{figure}[t] 
	\centering
	\includegraphics[width=8.5cm]{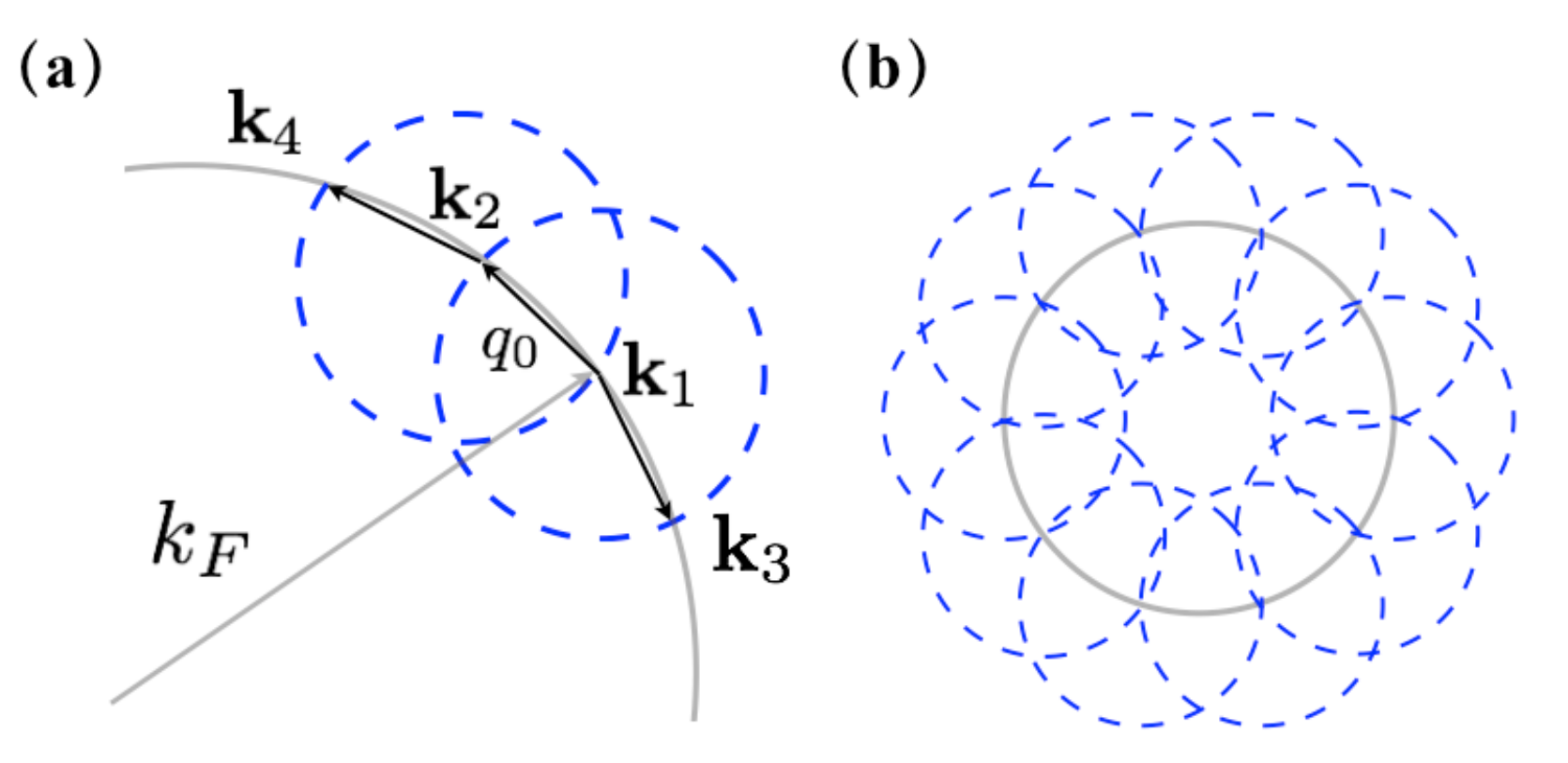}
	\caption{(Color online.) (a) The coupling between the (dashed blue) 
	 CBC and the (solid gray) Fermi surface. $k_F$ is the Fermi momentum, 
	and $q_0$ is the radius of the boson contour.
	(b) An example of commensurate coupling between the Fermi surface
	and the CBC. Here ${m=1,n=10}$ according to Eq.~\eqref{eq3}.}
	\label{fig2}
\end{figure}

Here we consider the coupling between the fermions on the Fermi surface and the bosonic modes. 
In the paramagnetic phase, the bosonic modes are fully gapped and do not
appear at low energies, thus the bosonic modes do not have qualitative 
impacts on the fermions on the Fermi surface.  
The fermion sector is a conventional Fermi liquid. 
We focus on the critical region where the CBC is relevant. 
Due to the CBC, each fermion mode (${\bm k}_1$ in Fig.~\ref{fig2}) 
on the Fermi surface is scattered to two other fermion modes (${\bm k}_2$ 
and ${\bm k}_3$ in Fig.~\ref{fig2}) on the Fermi surface.
As the fermion mode at ${\bm k}_2$ is further coupled
to the fermion mode at ${\bm k}_4$, eventually the whole Fermi surface
would be strongly coupled together except when the radius of the boson contour 
is commensurate with the Fermi momentum. The commensurability condition is 
established when multiple scatterings by the CBC bring the fermion back 
to the starting spot on the Fermi surface, {\sl i.e.}
\begin{equation}
\sin^{-1}(\frac{q_0}{2k_F}) = \frac{m\pi}{n} ,
\label{eq3}
\end{equation}
where $m$ and $n$ are both integers. This means, $n$-times scattering 
is able to bring the fermion to encircle the Fermi surface $m$ times and 
return to the starting point. 
For the commensurate case, 
the physics may look like the `hotspot theory' for the AFM quantum criticality
where the hotspots of the Fermi surface connected by the AFM ordering wavevectors become
critical, and here 
each fermion on the Fermi surface is coupled to the a finite number of fermion 
modes connected by the relevant critical boson modes. 
But there is a key difference. In the current case, all the fermions on the Fermi surface 
are critical, while there exist a finite number of critical fermions for the AFM quantum criticality. 

For more generic cases, such a commensurability condition in Eq.~\eqref{eq3} can not be 
satisfied. All the modes on the Fermi surface are eventually coupled together 
by the CBC which amounts to a `hot Fermi surface'. 
The presence of infinite many critical bosons on the CBC
introduce considerable complexity for the fermion-boson coupled system.
On one hand, the infinite many critical boson modes on the CBC resemble the FM quantum criticality;
On the other hand, the magnitude of the critical boson modes is finite and identical 
which resembles the AFM quantum criticality.
In this sense, we expect that the properties of both FM and AFM quantum criticalities are present
in the fermion-boson coupled system with the CBC in distinct quantum critical regimes.
It is extremely challenging to develop a unified theoretical framework to incorporate both FM and AFM quantum criticalities.
Starting from the next section, we make an attempt to decipher 
the complicated problem by considering the FM or AFM quantum criticality
in respective limits and relate the universal properties at the critical point to the transport phenomenon.
In a previous study on 3D counterpart of the fermion-boson coupled problem\cite{Zhang2023},
we are interested in the quantum critical regime mediated by the AFM quantum criticality;
for the present study, we explore the quantum critical regime with FM quantum criticality.
We hope these complementary studies unfold the complexity encoded in the quantum criticality
associated with infinite critical bosons and drawn more attention to this intriguing problem.


\subsection{Boson polarization function}
\label{sec:Pi}

For the 2D fermion-boson coupled system in the presence of the CBC, 
distinct low-energy dynamics can emerge for the critical bosons 
provided the ratio of the CBC and Fermi surface radius is in the respective limits.
We demonstrate this point by evaluating the boson polarization function 
due to particle-hole excitation around the Fermi surface.
We show that in the small boson radius limit $q_0/k_F \ll 1$,
the system mimics a FM quantum criticality with the dynamical critical exponent ${z=3}$.
In a previous study~\cite{Zhang2023}, 
we explore the consequence of fermions interacting with a commensurate boson radius
which amounts to the $z=2$ dynamical critical exponent. 
In this study, we consider the $z=3$ quantum criticality
and explain the crossover region between the two different types of quantum critical regimes.

The boson polarization function is the renormalized boson self-energy correction at the one-loop level
as illustrated by the Feynman diagram in Fig.~\ref{fig3}(a), which reads
\begin{equation}
\delta\Pi({\bm q},i\omega_l) = g^2 \int \frac{d\epsilon_n d^2{\bm k}}{(2\pi)^3} \frac{1}{i\epsilon_n -\xi_{\bm k}} 
\frac{1}{i(\epsilon_n+\omega_l) -\xi_{{\bm k}+{\bm q}}}. 
\end{equation}
${\xi_{\bm k} = {{\bm k}^2}/{(2m)}-\epsilon_F}$ is the parabolic dispersion around the Fermi level 
$\epsilon_F$. The kinematics of the Fermi surface and the CBC and 
their scattering configuration plays an important role in low-energy dynamics.
We adopt the so-called global coordinate~\cite{Shankar1994} for the fermion momentums
where the Fermi momentums are parametrized by the surface radius $k_F$ 
and a unit angular vector $\hat{k}_F$.
A given point near the Fermi surface is uniquely labeled by the nearest Fermi point
and the radial small momentum variation 
\begin{equation}
{\bm k}= \hat{k}_F(k_F + \delta k).
\label{k_decomp}
\end{equation}
The cubic form of the Yukawa interaction indicates that the boson momentum 
is the fermions momentum difference during the scattering.
The boson momentum near the CBC can be written as (see Fig.~\ref{fig2})
\begin{equation}
{\bm q} = {\bm q}_0 + \delta {\bm q}, \ \ \ \delta {\bm q} = \hat{q}_0 \delta q_\parallel + \delta {\bm q}_\perp.
\label{q_decomp}
\end{equation}

In a small boson radius limit 
\begin{equation}
q_0/k_F \ll 1,
\label{small_q0}
\end{equation} the fermion on the Fermi surface
is only scattered to nearby regions and the Fermi momentum is approximately
perpendicular to the bosonic one, {\sl i.e.} $|{\hat k}_F \cdot {\hat q}_0| \simeq \mathcal{O} ( q_0/k_F )$.
The fermion dispersions can linearize approximately as
\begin{eqnarray}
 \xi_{{\bm k}} &\simeq & 
 \frac{{\bm k}_F \cdot \delta {\bm k}}{m},
 \\ 
 \xi_{{\bm k}+{\bm q}} & \simeq & 
 \frac{ {\bm k}_F \cdot \big[ \delta {\bm k}+ ({\bm q}_0 +\delta {\bm q})\big]}{m}
 \end{eqnarray}
The boson polarization function is evaluated in the static limit $|\omega_l|\ll v_F \delta q$,
which yields
\begin{eqnarray}
  \delta\Pi(\delta{\bm q},i\omega_l) 
  &=& g^2 \int \frac{d\epsilon_n d(\delta k) k_F d\hat{k}_F}{(2\pi)^4}  \frac{1}{i\epsilon_n -v_F \delta k} 
\nonumber \\
&& \times\frac{1}{i(\epsilon_n+\omega_l) -v_F(\delta k +  {\hat k}_F \cdot \delta {\bm q} + q_0^2/k_F)  }
\nonumber \\
&= & i  \frac{g^2 k_F}{v_F} \int \frac{d\epsilon_n d\hat{k}_F}{(2\pi)^3}
\frac{{\rm sgn}(\epsilon_n) - {\rm sgn}(\epsilon_n+\omega_l)}{v_F {\hat k}_F \cdot \delta {\bm q} + q_0^2/m - i \omega_l } 
\nonumber \\
&\simeq & \frac{g^2 k_F}{4\pi^2 v_F}  \frac{\omega_l {\rm sgn}(\omega_l)}{ \sqrt{ (v_F \delta q)^2- (q_0^2/m)^2 }}.
\end{eqnarray}
We consider an intermediate momentum regime
\begin{equation}
  q_0(q_0/k_F) \ll \delta q < q_0.
\label{inter}
\end{equation}
This is a feasible regime because $q_0/k_F\ll 1$. 
This regime recovers the Landau damping term
that amounts to the dynamic exponent $z=3$ 
\begin{eqnarray}
 & & \delta\Pi(\delta {\bm q},i\omega_l) = \frac{|\omega_l|}{{\it \Gamma}_q}, 
 \\
 && {\it \Gamma}_q \equiv {\it \Gamma} (\delta q), \   {\it \Gamma}=  \frac{4\pi^2v_F^2}{g^2 k_F}. 
\label{deltaPi}
\end{eqnarray}
In a small momentum regime $\delta q < q_0(q_0/k_F)$, we obtain 
the Gilbert damping term $\sim |\omega_l|/{\it \Gamma}_{q_0}$
with ${\it \Gamma}_{q_0} =  \frac{g^2 m^2}{4\pi^2 q_0^2}$.

In addition, we check that when the small boson radius condition[Eq.(\ref{small_q0})] is violated,
the Gilbert damping with $z=2$ is obtained irrespective of the momentum regimes.
This is due to different ways of linearizing the fermionic dispersion around Fermi surface.
For a finite boson radius ${q_0 \lesssim k_F}$, a given Fermi point ${\bm k}_F$ 
in the vicinity of the Fermi surface is clearly scattered to a distinct point 
${\bm k}_F^\prime = {\bm k}_F + {\bm q}_0$ which yields
\begin{eqnarray}
&&  \xi_{{\bm k}} \simeq  {\bm k}_F \cdot \delta {\bm k}/m, \\
&& \xi_{{\bm k}+{\bm q}} \simeq {\bm k}_F^\prime \cdot ( \delta {\bm k} + \delta {\bm q})/m. 
 \end{eqnarray} 
The 2D momentum integral can be carried out in terms of two new variables: 
$k_1 = {\bm k}_F/m \cdot \delta {\bm k}, k_2 = {\bm k}_F^\prime/m \cdot \delta {\bm k}$.
The boson polarization function is calculated as,
\begin{eqnarray}
&& \delta\Pi(\delta {\bm q},i\omega_l) 
\nonumber \\
& &= \frac{g^2 }{|{\bm k}_F \times {\bm k}_F^\prime|/m^2} \int \frac{d\epsilon_n dk_1 dk_2 }{(2\pi)^3} \frac{1}{i\epsilon_n - k_1} 
\frac{1}{i(\epsilon_n+\omega_l) -k_2 }
\nonumber  \\
&& = \frac{g^2 }{|{\bm k}_F \times {\bm q}_0|/m^2}\int \frac{dk_1 dk_2}{(2\pi)^2} \frac{{\rm sgn}(k_2)-{\rm sgn}(k_1)}{k_1-k_2 - i\omega_l}
\nonumber \\
&& \simeq -\frac{g^2}{4\pi k_F q_0/m^2 } |\omega_l|. 
\end{eqnarray}

The two different damping terms affect the low-energy physics 
of the fermion-boson coupled system in the distinct fashions.
We have explored the NFL behaviors induced 
by the infinite many critical boson modes with ${z=2}$ 
in a previous study for the three-dimensional case~\cite{Zhang2023}.
In the present study, we make use of the Landau damping term obtained in the static limit
and study the NFL on the heterostructure interfaces induced by the magnetic fluctuations
that are described by an infinite number of critical boson modes with the university class 
${z=3, d=2}$. We note that developing a unified theoretical framework to incorporate both regimes
is quite difficult. We hope that the attempt here would raise the interest
in investigating the quantum criticality induced by infinitely many critical bosons
and its associated non-Fermi liquid behaviors when coupled to the fermionic degree of freedom.


\section{Magnetic fluctuations with critical boson contour}
\label{sec3}

In this section, we address the magnetic fluctuations due to the critical 
boson contour. The Ginzburg-Landau theory describes a 
continuous magnetic transition at zero temperature upon 
the variation of the tuning parameter $r$. Near the quantum 
critical point, the wild quantum fluctuation induces a non-Fermi liquid
behavior at low yet finite temperatures. {\emph J. Hertz} pointed 
out that the dynamical and static properties are intertwined 
and thus should be treated on equal footing~\cite{Hertz1976}.
A distinct universality class can be identified according to 
the dynamic exponent that is determined by the coupling
between the magnetic order parameter and the fermionic quasi-particles.
The bare term receives a self-energy correction
taking the form of the Landau damping term as derived in Eq.(\ref{deltaPi}).
The Landau damping term overwrites the original dynamic term in the low-energy limit
and the tangential component is introduced which turns out to be important for the low-energy and
low-temperature fluctuations.

In order to analyze the phenomena related to the novel boson fluctuations, 
we first carry out a scaling analysis for physical parameters and field operators on the tree level.
The validity of the Hertz-Millis method is examined for 
the university class $d=2,z=3$ with the presence of a critical boson contour.
Then, we implement a self-consistent renormalization (SCR) developed by \emph{T. Moriya}
to incorporate the effect of the interaction on the one-loop level.
The SCR scheme is built on the Hertz-Millis model
and is used to tame the wild fluctuation around QCP.



\subsection{Magnetic correlation}

We have shown that the low-energy physics of the interfacial magnetism
is dominated by the CBC, whose finite radius is guaranteed by the DM interaction.
The stability of the metallic Bose state was proposed theoretically 
a long time ago~\cite{Das1999,Phillips2003,Paramekanti2002,Motrunich2007},
which remains elusive until recent experimental observation 
in high-${T_c}$ superconductor films\cite{Yang2019,Phillips2019}
Recently, a fascinating Bose metallic phase with the CBC is proposed 
by the name ``Bose Luttinger liquid''\cite{Sur2019,Lake2021}.
The CBC plays a similar role as the Fermi surface in a Fermi liquid.
The stability of this unconventional critical phase is elucidated in terms of a patch theory\cite{Sur2019}.
The momentum-space annulus around the CBC is decomposed into patches,
and the inter-patch interaction is shown to be negligible in a certain limit.
The low-energy dispersion on each patch depends solely 
on the parallel component $\delta q_\parallel$.
This quasi-one-dimensional nature of the critical fluctuation 
is captured by a multi-dimensional bosonization scheme,
or equivalently the coupled wire construction\cite{Emery2000,Vishwanath2001,Yang2001,Yang2017,Zhang2017}.
These methods are known to be effective in studying competing orders above one dimension,
yielding a sliding Luttinger liquid description.
In this study, we are not concerned about the stability 
of the critical phase of bosonic matter which has been addressed extensively. 
Rather, we assume the interfacial magnetic phase transition stabilizes the CBC
and investigate the physical consequences in the quantum critical regime
for the magnetic sector as well as the electronic sector.


The structure in the boson correlation function sheds
light on the existence of the CBC. We evaluate the correlation function 
in the momentum and frequency space at the quadratic level.
The correlation function is given by the susceptibility 
matrix $M_{\mu\nu} ({{\bm q}, i\omega_l})$
\begin{eqnarray}
&& \big\langle \phi_{\mu} ({\bm q}, i \omega_l) 
\phi_{\nu} ({\bm q}', i \omega_{l'}) 
\big\rangle_{{\cal S}^{(2)}_{\rm B}} \nonumber \\
&& \quad\quad \quad\quad \quad\quad \quad
= M_{\mu\nu} ({{\bm q}, i\omega_l}) 
\,
\delta_{{\bm q} ,-{\bm q}'}
\,
\delta_{i \omega_l ,-i \omega_{l'}} ,
\end{eqnarray}
where the subindex ${{{\cal S}_{\rm B}}^{(2)}}$ refers to the 
statistical average over the quadratic action ${{{\cal S}_{\rm B}}^{(2)}}$.
The susceptibility matrix is an inverse of the polarization function
that reads
\begin{eqnarray}
&&  M({\bm q}, i\omega_l)
\nonumber \\
&& \,\,\, = \frac{1}{{\rm Det} [\Pi]}
 \left[\begin{array}{ccc}
f^2-D^2 q_y^2 & -D^2q_xq_y & -i f Dq_x\\
-D^2q_xq_y & f^2-D^2 q_x^2 & i f Dq_y \\
i f Dq_x & -i f Dq_y & f^2 \\
 \end{array}\right],
 \label{M_func}
\end{eqnarray}
where the trace of the polarization function reads 
${\rm Det}[\Pi ] =  3f^2-D^2q^2$.
The trace of Eq.~(\ref{M_func}) corresponds to the angular 
averaged spectrum at the criticality,
\begin{eqnarray}
{\cal F}({\bm q},\omega) &\sim &{\rm Tr}\big[ M({\bm q},\omega)\big] \nonumber \\
&=& \frac{1}{f-Dq} +\frac{1}{ f}+ \frac{1}{ f+Dq}.
\end{eqnarray}
 We notice that the first term conceives a divergent contribution from the CBC.
Therefore, we single out the divergent contribution at the criticality 
and the spectrum displays a divergent behavior around the CBC,
\begin{equation}
{\cal F}({\bm q},\omega) \sim \frac{(\omega/{\it \Gamma}_{q})}{J^2(q-q_0)^{4}+(\omega/{\it \Gamma}_{q})^2}.
\label{Spec_omega}
\end{equation}
In the static limit ${\omega/{\it \Gamma}_{q}\rightarrow 0}$,  
the expression of the spectrum in Eq.~(\ref{Spec_omega})\
corresponds to the Dirac delta function divergence in the 
momentum space for the approximation of the quadratic action.
 
\begin{figure}[t] 
	\centering
	\includegraphics[width=8cm]{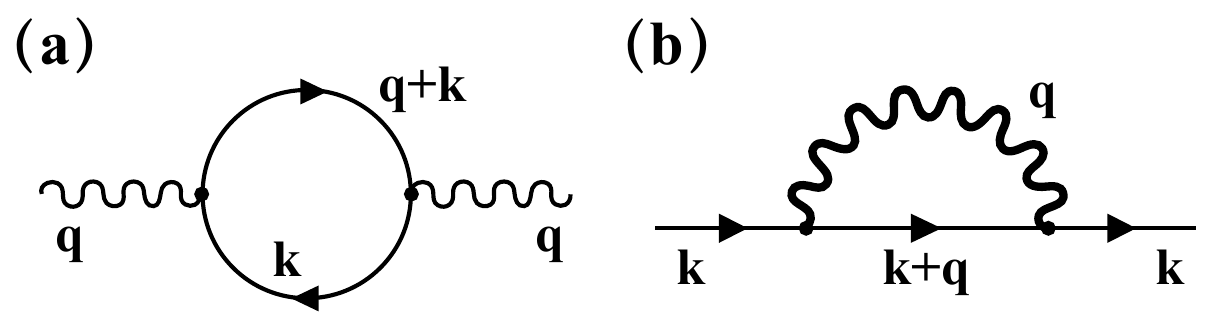}
	\caption{(a) The fermion bubble induced boson dynamics.  
	(b) The renormalized fermion propagator from the renormalized 
	boson correlator. The light and bold curly lines represent the 
	bare and renormalized boson correlators, respectively. 
	}
	\label{fig3}
\end{figure}

\subsection{Scaling analysis}

At the mean-field level, the magnetic correlation function gives a clear evidence
for the existence of the CBC which can be observed by spectroscopic detection.
The thermodynamic and transport measurement involves finer details
that go beyond the mean-field method, which requires properly taming the wild fluctuation.
We employ the SCR method to include the boson interactions in a perturbative manner.
The validity of SCR is built on the foundation of the Hertz-Millis method:
the bosonic sector has a Gaussian fixed point and the critical phenomenon 
is described by a Ginzburg-Landau expansion [see Eq.~\eqref{phi_4}].
Moreover, the Gaussian fixed point is immune to introducing fermion-boson coupling.


We justify the applicability of the Hertz-Millis method by carrying out 
the scaling analysis at the tree-level renormalization.
The scaling dimensions of the fermion and boson momentum variations
are determined by the low-energy scattering configuration illustrated in Fig.~\ref{fig4}.
We adopted the global coordinate to uniquely label the momentum variation around the Fermi surface,
namely only the radial component has a finite scaling dimension ${\rm dim}[\delta k]=1$
while the tangential component is treated as an angular variable with zero scaling dimension.
For a given Fermi momentum point, the boson momentum can be 
decomposed into components as in Eq.~\eqref{q_decomp}.
The radial and tangential components of boson momentums have scaling dimensions:
${{\rm dim}[\delta q_\parallel]=1, \ {\rm dim}[\delta q_\perp]=\gamma}$
where the value of $\gamma\in [\frac{1}{2},1]$ reflects the nature of curved boson contour~\cite{Nayak1994}.
We further take the convention that the fermionic and bosonic sectors have the same dynamic exponent 
${\rm dim}[\epsilon_n]={\rm dim}[\omega_l] = z$.

By requiring that the Gaussian parts of the bosonic and fermionic theories are invariant under renormalization,
we obtain scaling dimensions for various field operators and interactions [see detailed derivation in Appendix.~B].
The four-fermion interaction is irrelevant ensuring the Fermi liquid fixed point.
The quartic boson interaction $u$, with a scaling dimension ${\rm dim}[u]=3-z-(d-1)\gamma$,
as well as all higher-order interactions are all irrelevant for the university class $d=2,z=3$.
In addition, the Yukawa coupling constant is also irrelevant ${\rm dim}[g]=-\big[z-1+(d-3)\gamma\big]/2$.
These results are independent of the choice of $\gamma$ brought by the boson curvature effect.
In summary, the Ginzburg-Landau action [in Eq.(\ref{S_B})] is
a faithful expansion for the critical bosons which amounts to a Gaussian fixed point.
The Yukawa coupling can be therefore treated perturbatively.
and the Gaussian fixed point is stable against the fermion-boson coupling.

\subsection{Self-consistent renormalization}

In the vicinity of the FM or AFM  
quantum critical points, the fluctuation is governed by the ordering wavevector.
In contrast, in the presence of the CBC, the pattern of the fluctuation
is completely altered.  The low-temperature fluctuation would be 
governed by the CBC. To understand the novel features of the 
fluctuation in the critical properties, we adopt the self-consistent 
renormalization (SCR)~\cite{Moriya1973,Moriya2006}. 
This approach was found to be quite effective 
in the study of ferromagnetic and antiferromagnetic criticalities,
and gave consistent results about physical properties with the
experimental measurements. The spirit of the SCR approach 
is essentially variational. It is to find the most appropriate
variational free energy at the quadratic level to replace the 
original free energy that contains the quartic interaction. 
This approach amounts to decoupling the quartic interaction
and incorporating the interaction with a self-consistent
Born approximation.  
Technically, it corresponds to conducting a renormalized 
one-loop approximation for the scattering of the order parameters.
The renormalization process endows the non-thermal parameter 
$r$ with a temperature dependence ${r(T)=r_c +\delta(T)}$ after 
considering the effect of thermal fluctuation and correlation.
Here, $\delta(T)$ measures the relative distance towards the critical point
and is associated with the inverse square of the correlation length
which diverges from approaching the critical point.
The temperature dependence of $\delta$ (or equivalently $r$)
plays an important role in determining low-temperature thermodynamics.

To find the appropriate action and determine the scaling law of $\delta(T)$, 
one relies on Feynman's variational approach to optimize the free energy. 
The proposed variational free energy for the bosonic sector is
of the following form,  
\begin{equation}
F ({\tilde r}) \equiv \tilde{F} ({\tilde r})
 + \frac{1}{\beta} \langle {\cal S}_{\rm B} 
 - \tilde{\cal S}_{\rm B}  \rangle_{\tilde{\cal S}_{\rm B}},
\label{Var_F}
\end{equation}
where the quadratic part of the variational free energy is
\begin{equation}
\tilde{F} (\tilde{r})  \equiv -\beta^{-1} \ln {\rm Tr} \big(e^{-\tilde{\cal S}_{\rm B}}\big).
\end{equation}
The variational free energy ${\tilde{\cal S}_{\rm B}} (\tilde{r})$
is defined by introducing a variational parameter $\tilde{r}$, 
which remains undetermined, in exchange for the control parameter 
$r$ in Eq.~\eqref{S_B} and Eq.~\eqref{boson_Pi}. 
 We then establish the saddle point equation with respect to $\tilde{r}$.
The SCR equation ${\partial_{\tilde{r}} F(\tilde{r}) =0 }$ has an expression
\begin{equation}
\tilde{r} = r + \frac{cu}{\beta V} \sum_{{\bm q}, i\omega_l} {\rm Tr}\big[M({\bm q}, i\omega_l)\big],
\label{SCR_TrM}
\end{equation}
where $c$ is a non-universal number. 
We solve the self-consistent equation at low temperatures 
and provide the technical details in a step-by-step manner in the appendix.
The solution in favor of $\delta(T)$ 
exhibits a simple scaling law
\begin{equation}
\delta(T) = T^{\alpha}, \quad \alpha=\frac{5}{9} .
\label{delta_T}
\end{equation}
As promised, the non-thermal parameter acquires a $T$-dependence 
${r(T)= r_c + \delta(T)}$ as a result of the SCR procedure.
The exponent is much smaller than 
the values for 2D FM (${\alpha=4/3}$) and 
AFM (${\alpha=3/2}$) criticalities.
Moreover, this value is different from the one 
obtained for the 3D itinerant quantum magnets in the presence 
of a critical boson sphere~\cite{Zhang2023}. 
On one hand, the fluctuating modes near the CBC are dispersive along the radial direction
which is illustrated in Fig.~\ref{fig4}.
Namely, the system is of quasi-one-dimensional
nature regardless of the dimensionality of the critical boson manifold.
On the other hand, the boson polarization in the long-distance limit
takes the form of Landau damping term with $z=3$ that contains the tangential components.
The Landau damping term completely alters the low-energy dynamics
of the fermion-boson coupled system and leads to a peculiar power-law scaling for $\delta(T)$.

In the next section, we show that the structure of the CBC 
makes a huge difference when considering the scattering process with the fermions. 
The parameter $\delta(T)$ contributes to the thermodynamic properties 
of itinerant electrons in an important way.

\subsection{Thermal crossover to FM criticality}

As we have previously emphasized, the interfacial DM interaction would
convert the interfacial ferromagnet into 
a frustrated magnet where there exists a degenerate momentum
contour for the system to select the magnetic order. 
For the continuum model that was discussed above, the 
fluctuation near the CBC dominates the low-temperature 
and the-energy physics. As the  
temperature further increases, the thermal fluctuation
 overcomes the curvature of the CBC
 such that the temperature is comparable to 
 the energy difference between the ${\it \Gamma}$ point and the CBC.
 This crossover temperature is estimated to
 be of the order of $\mathcal{O} (D^2/J)$. 
 Above this crossover temperature and in the critical regime,
 compared to the large thermal fluctuation, the 
 CBC behaves very much like a point. Therefore, the system 
 would experience the crossover to the ferromagnetic criticality (see Fig.~\ref{pdiagram}).
 This crossover would also show up in the transport properties of the
 itinerant electrons.

\section{Non-Fermi liquid behaviors}
\label{sec4}

In the conventional non-Fermi liquid near the 
FM or AFM criticality,
the low-energy theory of the bosonic sector 
comprises discrete momentum points. Here, 
we encounter an infinite number of bosonic modes 
on a contour that become critical simultaneously. 
The quasiparticles near the Fermi energy are scattered 
by the CBC, resulting in a novel type of non-Fermi liquid 
in 2D. This is depicted in the quantum critical regime (painted in blue) 
in the $(T,r)$-phase diagram in Fig.~\ref{pdiagram}.
In this section, we first evaluate the frequency dependence of the fermionic self-energy correction at zero temperature,
which indicates that the quasiparticles no longer exist.
Then, we evaluate the temperature dependence of the fermionic self-energy with zero external frequency
and predict the non-Ferm liquid behaviors in transport experimental observables.

\begin{figure}[t]
\includegraphics[width=7cm]{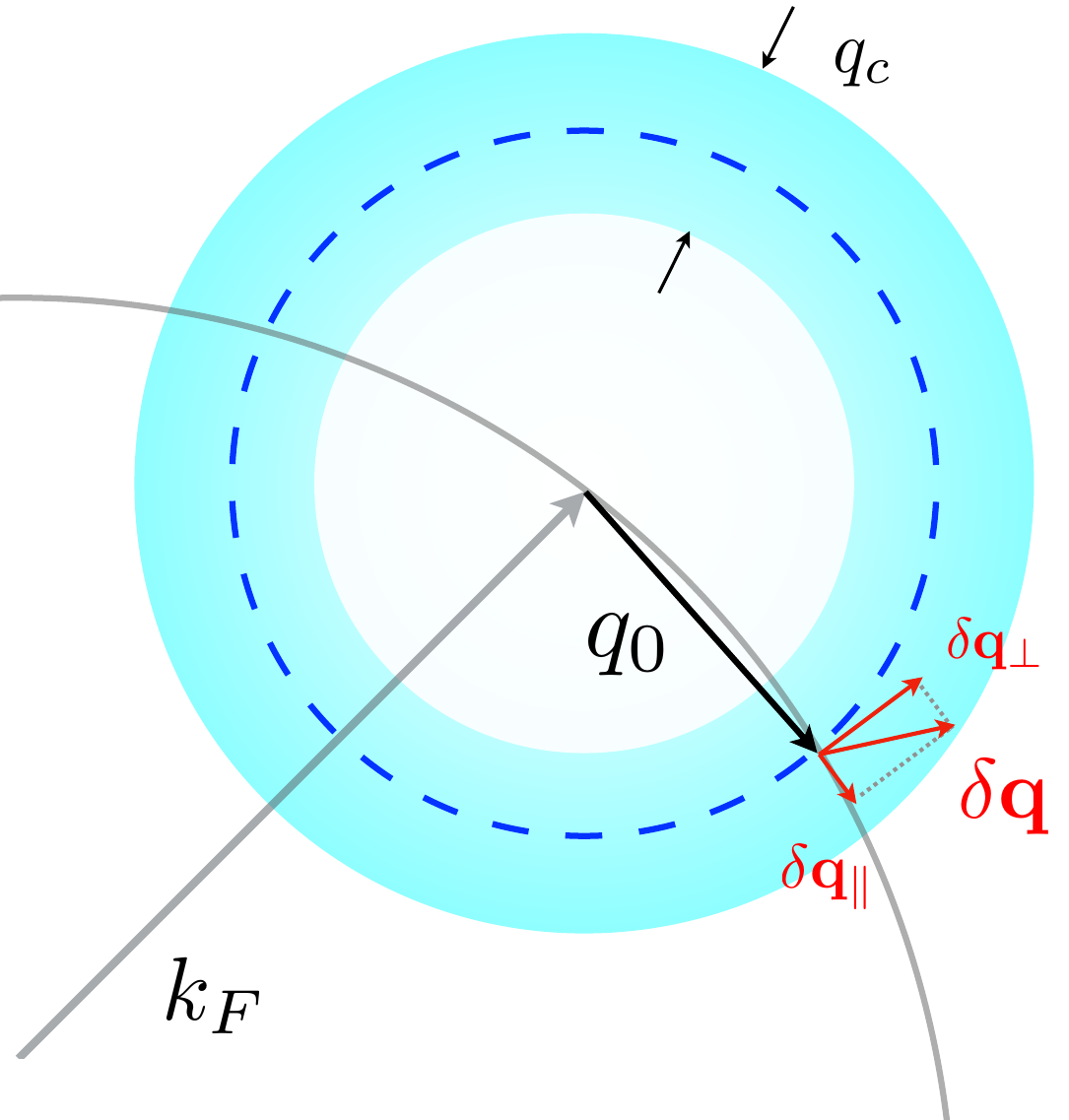}
\caption{(Color online.) The scattering process of the 
fermions on the Fermi surface by the bosons near the CBC. 
The (gray) curve stands for the Fermi surface
that is coupled to the (dashed) CBC of the bosonic sector. 
The (red) vectors are the small momentum variation from the CBC. 
The boson momenta ${\bm q}_0$ and $\delta {\bm q}$ are exaggerated 
for clarity. The boson momentum deviation $\delta {\bm q}$ is 
decomposed into the parallel (perpendicular) component 
$\delta {\bm q}_\parallel$ ($\delta {\bm q}_\perp$). 
The parallel direction is aligned to the vector ${\hat q}_0$. }
\label{fig4}
\end{figure}

\subsection{Zero temperature}

The existence of quasiparticles in fermionic sector
has to be re-examined in the presence of scattering process introduced by the CBC.
Since the fermion-boson cubic interaction flows to a weak coupling limit,
we can estimate the fate of quasiparticles perturbatively in terms of Yukawa coupling $g$.
We evaluate the real frequency dependence of the fermionic self-energy 
at zero temperature $\Sigma({\bm k}_F,\omega;T=0)$.
Specifically, the fermion self-energy correction induced by the CBC is given by,
\begin{equation}
\begin{aligned}
\Sigma({\bm k}_F,i\epsilon_n;T=0) = & g^2 \int \frac{d^2 {\bm q}}{(2\pi)^2}\frac{d\omega_l}{2\pi} \frac{1}{(q-q_0)^2+ |\omega_l|/\Gamma_q}   \\
& \times \frac{1}{i(\epsilon_n+\omega_l)- \xi_{{\bm k}_F+{\bm q}}}. \\
\end{aligned}
\label{Sig_T0}
\end{equation}

The momentums are decomposed in the same way as in Eq.(\ref{k_decomp},\ref{q_decomp}).
The integrand involves all components of $\delta {\bm q}$ 
which is due to the special scattering configuration facilitated by the CBC.
The integral can be carried out by using a polar coordinate $\delta {\bm q}= \delta q(\cos\phi,\sin\phi)$,
which reads
\begin{equation}
\begin{aligned}
\Sigma &({\bm k}_F,i\epsilon_n;T=0)  = \frac{g^2}{4\pi^2}  \int\frac{d\omega_l}{2\pi} \int_0^{+\infty} d(\delta q) \delta q \int_0^{2\pi} d\phi \\
& \times \frac{1}{ (\delta q)^2 \sin^2\phi+ |\omega_l|/(\Gamma \delta q)}
\frac{1}{i(\epsilon_n+\omega_l)- v_F \delta q \cos\phi}.\\
\end{aligned}
\end{equation}
The leading contribution in the low-energy limit is calculated in Appendix.~C
which takes a form $\Sigma ({\bm k}_F,i\epsilon_n;T=0) \sim i {\rm sgn}(\epsilon_n) |\epsilon_n|^{2/3}$.
By performing an analytic continuation to the real frequency $i\epsilon_n \rightarrow \omega+ i 0^+$,
we obtain 
\begin{equation}
{\rm Im}\Sigma ({\bm k}_F,\omega;0) \sim {\rm sgn}(\omega) |\omega|^{2/3}.
\end{equation}
The broadening of the quasiparticle peaks is significantly enlarged compared with the Fermi liquids $\sim \omega^2$.
We further use the Kramer-Kronig relation to obtain the real part of the self-energy
which leads to the definition quantity for Fermi liquid: 
the quasiparticle weight $Z({\bm k}_F, \omega) = \big[ 1- \partial_\omega {\rm Re}\Sigma ({\bm k}_F,\omega;0) \big]^{-1}$.
In the low-energy limit $\omega\rightarrow 0$, one can readily see that the quasiparticle weight vanishes.
All the evidences indicate the destroy of quasiparticles and emergence of non-Fermi liquids.

We note that despite the presence of the CBC, the fermionic self-energy correction at zero temperature
is exactly same as the Ising nematic QCP (with $d=2,\ z=3$) induced NFLs\cite{Metlitski2010}.
The Hertz-Millis scheme relies on integrating out the gapless fermions first
which generates low-energy dynamical term for bosons;
and, the renormalized bosons are used to further evaluate the fermionic self-energy.
With the fermionic self-energy correction, the self-consistency of the Hertz-Millis scheme
has been checked in literatures\cite{Metzner2006,Metlitski2010} 
by re-inserting the renormalized fermionic Green function into the bubble diagram in Fig.~\ref{fig3}(a).
The one-loop expressions in fact satisfy the Eliashberg-type equations
and are therefore self-consistent solutions.

The frequency dependence can \emph{not} be extended to finite temperatures
by simply assuming the $\omega/T$ scaling behavior.
The $T$-dependence for the self-energy has multiple sources due to fluctuations around the CBC
which reflects in the peculiar exponent derived by SCR method [see Eq.(\ref{delta_T})].
In the next section, we carefully take all the contributions into account
and predict non-Fermi liquid behaviors at finite temperatures.

\subsection{Finite temperature}

In the vicinity of the critical point, the non-Fermi liquid behavior manifests 
as a characteristic temperature scaling in the thermodynamic properties.
In particular, we focus on the resistivity that describes the transport property.
To this end, we evaluate the (renormalized) fermionic self-energy 
making use of the renormalized boson correlation function. 
The one-loop renormalization sketched in Fig.~\ref{fig3}(b) 
is explicitly expressed as
\begin{eqnarray}
&& \Sigma ( {\bm k}, i\epsilon_n )
\nonumber \\ 
  && \quad \simeq 
 \frac{g^2}{\beta V} \sum_{{\bm q},i\omega_l}
 G_0({\bm k}+{\bm q},i\epsilon_n+i\omega_l) 
 \text{Tr} [ M({\bm q},i\omega_l) ] ,
\end{eqnarray}
where ${\epsilon_n\equiv(2n+1)\pi/\beta}$ ${(n \in {\mathbb Z})}$ 
is the Matsubara frequency for the fermions, and
${G_0({\bm k},i\epsilon_n)= (i\epsilon_n-\xi_{\bm k})^{-1}}$
is the bare Green's function of the fermions. 
We follow the argument in the paragraphs near Eq.~(\ref{M_func})
and single out the term that contains the CBC 
in the boson susceptibility. After transforming to the 
dispersive representation and performing an analytic 
continuation ${i\epsilon_n \rightarrow \omega+i 0^+}$,
we express the imaginary part of the fermion self-energy 
as
\begin{equation}
\begin{aligned}
{\rm Im} \Sigma^{\rm R}({\bm k},\omega;T) 
= & {-\frac{g^2}{V} } \sum_{\bm q} \int_{-{\it \Gamma}_q}^{+{\it \Gamma}_q} 
d\epsilon \big[ f_B(\epsilon) +f_F(\epsilon+\omega) \big] \\
& \times \delta(\omega+\epsilon-\xi_{\bm k+q}) 
\frac{\epsilon/{\it \Gamma}_q}{{\cal E}_q^2+(\epsilon/{\it \Gamma}_q)^2},  \\
\label{Im_Sigma_R}
\end{aligned}
\end{equation}

where we have used the relation ${|\epsilon|<{\it \Gamma}_q}$.
We are interested in the $\omega=0$ static component 
at low temperatures $|\epsilon|\ll T$.
Together with the zero temperature limit $|\epsilon|\ll T$,
we are interested in the regime set by a sequence of energy scales
\begin{equation}
|\epsilon| < {\it \Gamma}_q \ll T.
\label{T_cond}
\end{equation}
This condition imposes a temperature-dependent bound on
the magnitude of the momentum variation: $\delta q \le q_c = T/{\it \Gamma}$.
The following calculations are performed under the condition in Eq.(\ref{T_cond})
which yields
\begin{equation}
\begin{aligned}
-{\rm Im} \Sigma^{\rm R}({\bm k},0;T) 
= &  \frac{g^2}{V} \sum_{\bm q}  \frac{T/{\it \Gamma}_q}{{\cal E}_q^2+(\xi_{\bm k+q}/{\it \Gamma}_q)^2} , \\
 \label{Im_Sigma_Cond}
\end{aligned}
\end{equation}
where the integration involving the delta function generates a finite result 
provided a reasonable condition $v_F\le {\it \Gamma}^{-1}$ is satisfied with a small interacting strength $g$.

To properly extract the low-temperature behavior from
the imaginary part of the fermion self-energy, 
we keep track of all the temperature dependence 
in Eq.~(\ref{Im_Sigma_Cond}) that are conceived in:
i) the renormalized thermal parameter $\delta(T)$;
ii) the upper bound for the momentum integrals. 
The scattering processes that are relevant at the low temperatures
are the elastic scattering process where the exchanged 
energy is nearly zero ${\xi_{\bm k+q}\simeq 0}$.
In other words, the Fermi momenta on the Fermi surface 
${|{\bm k}| = k_F}$ are scattered back to the vicinity of the Fermi surface 
with ${|{\bm k}+{\bm q}_0| = k_F}$. We schematically show 
this scattering configuration in Fig.~\ref{fig4}.
The deep blue dashed line is the CBC and the light blue region 
represents the low-energy shell around the CBC.
Let us consider the low-energy scattering process
associated with the intersecting region between the Fermi surface and the CBC.
Here, we follow the same convention as the calculation of the boson polarization function in Sec.~\ref{sec:Pi}.
The fermionic and bosonic momentums are expanded in the manner
as explained in Eq.(\ref{k_decomp}) and Eq.(\ref{q_decomp}).
And, we consider the reasonable limiting cases with:
i) the static limit dictated in Eq.(\ref{inter});
ii) the small boson radius limit in Eq.(\ref{small_q0}).
Under these conditions, the directional vectors ${\hat k}_F, {\hat q}_0$
form the local basis of the 2D coordinate.
The bosonic momentum variation can be further decomposed 
into components that are parallel and perpendicular with respect to $\hat{q}_0$,
which reads
\begin{equation}
\delta {\bm q} = \delta {\bm q}_\perp + \delta {\bm q}_\parallel, \ \  
\delta {\bm q}_\parallel = \delta q_\parallel \hat{q}_0, \ \ \delta {\bm q}_\perp = \delta q \hat{k}_F. 
\end{equation}
The fermion dispersion function is expanded
with respect to small bosonic momentum fluctuations
\begin{equation}
\xi_{\bm k+q} =  \frac{({\bm k}+{\bm q}_0 + \delta {\bm q})^2}{2m} - \epsilon_F 
 \simeq  v_F \delta q_\perp,
\label{xi_dq}
\end{equation}
where the leading order term ${o(\delta q)}$ is retained.
Observing the $\delta q$-integral in Eq.(\ref{Im_Sigma_R}), 
we note that the dominant contribution comes from the region
${\delta +\frac{1}{2} (\delta q_{\parallel})^2 \sim \xi_{{\bm k}+{\bm q}}/\Gamma_q
\sim (v_F/\Gamma) (\delta q_\perp/\delta q) }$.
We can safely drop the term with ${\mathcal O}(\delta q_\parallel)^2$
and carry out the integration in the polar coordinate 
$(\delta {\bm q}_\parallel, \delta {\bm q}_\perp)= \delta q (\cos\phi,\sin\phi)$,
which leads to
\begin{equation}
\begin{aligned}
- {\rm Im} \Sigma^{\rm R}({\bm k},0;T) & =   g^2 \int_{0}^{q_c} (\delta q) d(\delta q) \frac{T}{\Gamma (\delta q)} \\
& \times \int_0^{2\pi}d\phi
\frac{1}{{\delta}^2+ \big[ (v_F/\Gamma) \cos\phi\big]^2} \sim  \frac{T q_c}{\delta}. \\
\end{aligned}
\label{q_perp}
\end{equation}
Substituting the expression of $\delta(T)$ into Eq.~(\ref{q_perp}),
we end up with a temperature dependence for the imaginary part
 of the fermion self-energy in the low-temperature limit
\begin{eqnarray}
- {\rm Im} \Sigma^{\rm R}(k_F,0;T)
&\sim & T^{2-\alpha}\sim T^{1.44}.  
\label{SM_Sig_T}
\end{eqnarray}

The imaginary part of the self-energy in Eq.~\eqref{SM_Sig_T} 
corresponds to the inverse lifetime function of the electron 
on the Fermi surface ${|{\bm k}|=k_F}$. According to the Drude formula, 
the electronic resistivity is proportional to the inverse mean lifetime,
which exhibits a temperature dependence 
\begin{equation}
\begin{aligned}
\rho(T)\sim \frac{1}{\tau}\sim  T^{1.44 }.
\end{aligned}
\end{equation}
This peculiar temperature scaling exponent of the resistivity
signatures a novel type of non-Fermi liquid behavior.
The power-law $\rho(T)$ can be compared 
with the FM and AFM criticalities in 2D, 
where the standard Hertz-Millis theory yields 
${\rho (T) \sim T^{4/3}}$ and ${\sim T}$ respectively~\cite{Stewart2001}.
We conclude by claiming that the existence of the CBC fundamentally changes 
the scattering process and generates a distinct power law for the resistivity at finite yet 
low temperatures.

\section{Discussion}
\label{sec5}



The external magnetic field has an important impact on the  
physical properties near a quantum critical point. 
The magnetic field is often used to tune the properties
of the quantum criticality. Here, an external magnetic field explicitly 
breaks the time-reversal symmetry and immediately 
modifies the universality class of the phase transition.  
The temporal quantum fluctuations of the order 
parameter (boson field) is crucial in determining 
the university class of the quantum phase transitions.   
Within the framework of spin fluctuation theory, 
the presence of a magnetic field introduces a new type   
of dynamics in addition to the Landau damping.
The external magnetic field induces a precession for the spin order parameter, 
and the Lagrangian in Eq.(\ref{phi_4}) would be
 supplemented with a dynamic precession term 
$\sim \int d^2{\bm x}\ {\bm B}\cdot i\big(\phi \times  \partial_\tau \phi\big)$~\cite{Fischer2005}.
For conventional antiferromagnetic criticality, however, 
the Landau damping and precession 
belongs to the same university class with ${z=2}$.
The interplay of the two dynamic processes 
in 3D merely leads to non-universal changes
in various thermodynamic quantities depending 
on the relative magnitudes~\cite{RMP2007,Fischer2005}.
In contrast, the precession of the spin order parameter
changes the nature of the 2D interfacial magnetic transition qualitatively.
In the strong out-of-plane field limit, the critical boson contour
is preserved and the dynamical critical exponent 
is converted to ${z=1}$, 
which is in the same class as conventional insulating magnets.
The result in this work no longer applies, and the impact 
on the itinerant electrons will be explored in future work. 
The in-plane magnetic field, however, would compete with the 
DM interaction, and the critical boson surface
for the minima would not be satisfied.


In the formulation of our theory, we have assumed a global 
U(1) symmetry and the interfacial DM
interaction respects this continuous U(1) symmetry. 
In reality, this symmetry can be weakly broken by various 
anisotropy in the system, and the continuous degenerate
contour will be lifted. As long as the temperature or the 
energy scale appears above the anisotropic energy scale,
the results of the non-Fermi liquid and critical boson
contour would hold. We thus vision that our results 
can apply to a large temperature/energy window. 
In addition, the contour or line degeneracy appears
constantly in frustrated magnetic systems in frustrated 
magnets like honeycomb lattice antiferromagnets~\cite{PhysRevB.81.214419,Yao_2021,Gang2020,Varney2012},
and can even be present in more exotic excitations like
spinons or magnetic monopoles in spin liquids~\cite{PhysRevB.94.205107}.
Therefore, in the itinerant frustrated 
materials, such non-Fermi liquid behaviors can potentially 
emerge. 
Finally, how the strong fluctuations of the quasi-long range ordered regime 
in Fig.~\ref{pdiagram} impact on the itinerant electrons
is an interesting topic on its own. This problem differs qualitatively 
from the coupling between the gapless Goldstone mode
and the itinerant electrons where there can lead to non-Fermi liquids~\cite{Zhang2023,Watanabe_2014}, 
and will be considered in future works.

It is worthwhile to comment on the case when the itinerant electrons
on the heterostructure interface are Dirac fermions or involve
discrete band touchings. This happens for example on the surface or
the termination of a three-dimensional topological insulator where
magnetism can occur due to the correlation. 
At the charge neutrality point, the Fermi surface is composed 
of discrete momentum points. Only if there exist boson modes
on the CBC that connect the momentum points, the low-energy
physics can be reduced to the coupling between these fermions 
and the relevant critical bosons. Thus, most often, the CBC
scatters the fermions at the band touching points to the gapped region 
in the reciprocal space.  Such processes are certainly not low-energy 
physics, and cannot provide a good starting point for the analysis
of low-energy physics. In fact, the optimization of the kinetic energy
for the fermions could possibly select certain magnetic ordering 
wavevectors from the boson contour in the ordered regime. 
This is quoted as ``fermion order by disorder''.

In summary, we have studied the itinerant quantum magnets
at the interface of the magnetic heterostructure. 
We propose a novel type of non-Fermi liquid
that is induced by infinite many critical boson modes
due to the existence of interfacial DM interaction.
We uncover the physical consequences of the corresponding
experimental features for the itinerant electrons and 
local spins in the spectroscopy and transport, respectively.
Our study has broadened the physical context
for the exotic bosonic critical phases of matter 
with continuously degenerate minima, which seem to become 
a focused, rapidly developing research field recently~\cite{Zhang2023,Yang2017,Lake2021,Tsvelik2021,Ku2021}.

\appendix

\section{Calculation of $\delta(T)$ at low temperatures}

We calculate the temperature dependence of $\delta(T)$
by solving the saddle point equation given in Eq.~\eqref{SCR_TrM}. 
The important contribution of the CBC can be singled out,
\begin{eqnarray}
\delta & \simeq & \, \delta_0 + \frac{cu}{\beta V}  \sum_{q, i\omega_l} \frac{1}{f-Dq}  
\nonumber  \\
& = &\, \delta_0 +  \frac{2cu}{\beta V} \sum_{q, i\omega_l>0} \frac{1}{{\cal E}_q+\omega_l/{\it \Gamma}_q} . 
\label{delta_def}
\end{eqnarray}
Instead of performing the summation over Matsubara frequency directly,
we invoke a dispersive representation by using the Kramer-Kronig relation,
\begin{equation}
 \frac{1}{f({\bm  q},i\omega_l)-Dq}
=  \frac{1}{\pi} \int_{-\infty}^{+\infty} d\epsilon \frac{\tilde{f}(\epsilon,{\bm  q})}{\epsilon-i\omega_l},
 \label{SM_KK_relation}
\end{equation}
where the Matsubara frequency is always positive.
The function $\tilde{f}(\epsilon,{\bm  q})$ is defined as 
an imaginary part of a retarded function
\begin{eqnarray}
\tilde{f}(\epsilon,{\bm  q}) &
\equiv & {\rm Im} \frac{1}{f({\bm  q},i\omega_l\rightarrow \epsilon+i\eta)-Dq}
\nonumber  \\
& = & \frac{\epsilon/{\it \Gamma}_q}{{\cal E}_q^2 + (\epsilon/{\it \Gamma}_q)^2} .
 \label{SM_g}
\end{eqnarray}
Owing to the fact that $\tilde{f}(\epsilon,{\bm  q})=-\tilde{f}(-\epsilon,{\bm  q})$,
we rewrite the integral as the following and carry out the $\omega_l$-summation
\begin{eqnarray}
&&\frac{1}{\beta}\sum_{\omega_l} \int_{-\infty}^{+\infty} d\epsilon \tilde{f}(\epsilon,{\bm  q}) \frac{1}{\epsilon-i\omega_l} \nonumber\\
&& =  \frac{1}{2}  \int_{-\infty}^{+\infty} d\epsilon \tilde{f}(\epsilon,{\bm  q}) \frac{1}{\beta}\sum_{\omega_l} \Big\{\frac{1}{\epsilon-i\omega_l} -\frac{1}{-\epsilon-i\omega_l} \Big\} \nonumber\\
& &=  \frac{1}{2}  \int_{-\infty}^{+\infty} d\epsilon \tilde{f}(\epsilon,{\bm  q}) \coth(\epsilon/2T). 
\label{SM_KK}
\end{eqnarray}
 
With the results from Eqs.~\eqref{SM_g} and \eqref{SM_KK},
we arrive at an expression 
\begin{equation}
\begin{aligned}
\delta = \delta_0 + {cu} \int\frac{d^{3}{\bm  q}}{(2\pi)^3}\int_0^{{\it \Gamma}_q}\frac{d\epsilon}{2\pi} 
\coth\frac{\epsilon}{2T} \frac{\epsilon/{\it \Gamma}_q}{{\cal E}_q^2+ (\epsilon/{\it \Gamma}_q)^2}.
\end{aligned}
\label{SM_delta_delta_0}
\end{equation}
At zero temperature, the above equation reduces to
\begin{equation}
\begin{aligned}
0= \delta_0 +cu \int\frac{d^{3}{\bm  q}}{(2\pi)^3}\int_0^{{\it \Gamma}_q}\frac{d\epsilon}{2\pi}  \frac{\epsilon/{\it \Gamma}_q}{(q-q_0)^4+ (\epsilon/{\it \Gamma}_q)^2}.
\end{aligned}
\end{equation}
We obtain an expression for $\delta_0$,
which is substituted into Eq.~\eqref{SM_delta_delta_0}.
We end up with an equation for $\delta$ 
\begin{eqnarray}
\delta &=&  cu \int\frac{d^{3}{\bm  q}}{(2\pi)^3}\int_0^{{\it \Gamma}_q}\frac{d\epsilon}{2\pi} 
\big(\coth\frac{\epsilon}{2T}-1\big) \frac{\epsilon/{\it \Gamma}_q}{{\cal E}_q^2+ (\epsilon/{\it \Gamma}_q)^2}
\nonumber \\
& & + \frac{cu}{2} \int\frac{d^{3}{\bm  q}}{(2\pi)^3}\int_0^{{\it \Gamma}_q}\frac{d\epsilon}{2\pi} 
\Big\{ \frac{\epsilon/{\it \Gamma}_q}{{\cal E}_q^2+ (\epsilon/{\it \Gamma}_q)^2} 
\nonumber \\
&& \quad\quad\quad\quad\quad -\frac{\epsilon/{\it \Gamma}_q}{(q-q_0)^4+ (\epsilon/{\it \Gamma}_q)^2} \Big\}.
\label{SM_delta_delta_0}
\end{eqnarray}
 
When solving this equation, it should be kept in mind that 
the contribution from small frequency $\epsilon$ 
and small momentum variation around critical boson surface 
$\delta {\bm  q}={\bm  q}-{\bm  q}_0$ are essential.\\

We consider the low-temperature regime
and the boson distribution function admits an expansion 
$\coth(\epsilon/2T) -1 = 2f_{\rm B}(\epsilon) \simeq 2T/\epsilon$.
Importantly, we assume as a prior that 
\begin{equation}
{\delta(T)\sim T^\alpha}, \ \ {\alpha <1}.
\label{delta_alpha}
\end{equation}

The two terms in Eq.~\eqref{SM_delta_delta_0} are evaluated separately
under a low-temperature approximation,
$|\epsilon|/T\ll 1$, which sets up an upper limit for the frequency integral.
The first term is denoted as $\delta_1(T)$ and is calculated as,
\begin{equation}
\begin{aligned}
\delta_1
\simeq & \frac{cu}{2}  \int \frac{d^2{\bm  q}}{(2\pi)^2} \frac{2T}{{\it \Gamma}_{q}}  \int_0^{T}   \frac{{d\epsilon}/{\pi}}{(\delta+r_{\rm c}-D q+q^2)^2+(\epsilon/{\it \Gamma}_{q})^2}   \\
= &  \frac{cu}{2}  \int \frac{d^2{\bm  q}}{(2\pi)^2} \frac{2T}{{\it \Gamma}_{q}} \frac{1}{|\delta+(q-q_0)^2|} \tan^{-1}\frac{T/{\it \Gamma}_{q}}{|\delta+(q-q_0)^2|}   \\
= & cu \int \frac{d(\delta {\bm q}_\parallel)}{(2\pi)}\int \frac{d(\delta {\bm q}_\perp)}{(2\pi)}  \frac{T/{\it \Gamma}}{\sqrt{(\delta q_\parallel)^2+(\delta q_\perp)^2}} \\
& \times  \frac{1}{\delta+(\delta q_\parallel)^2} \tan^{-1}\frac{(T/{\it \Gamma}) }{\sqrt{(\delta q_\parallel)^2+(\delta q_\perp)^2}\big[\delta+(\delta q_\parallel)^2 \big]} ,  \\
\label{SM_delta1}
\end{aligned}
\end{equation}
where we have invoked the relation ${r_{\rm c}=q_0^2= (D/2)^2}$.
As illustrated in supplementary Fig.~\ref{fig4}, 
the ${\bm  q}$-integral in the momentum shell transforms into
the momentum variation $\delta {\bm  q}_\parallel$-integral
which is cutoff at at a bound $\pm q_{\rm c}/2$; as well as an average over the CBC.
We make a transformation on the radial and tangential components
$x= \delta q_\parallel/(T/{\it \Gamma})^{1/3}, \ y= \delta q_\perp/(T/{\it \Gamma})^{1/3}$
and rewrite the integrals as
\begin{equation}
\begin{aligned}
\delta_1
= & \frac{cu}{4\pi^2} (T/{\it \Gamma})^{2/3} \int_0^{x_c} dx  \int_0^{x_c} dy  \frac{1}{\sqrt{x^2+y^2}(x_0^2+x^2)} \\
& \times  \tan^{-1}\frac{1}{\sqrt{x^2+y^2}(x_0^2 + x^2)}   \\
\simeq & \frac{cu}{4\pi^2} (T/{\it \Gamma})^{2/3} \int_{x_0}^{x_c} dx  \int_{x}^{x_c} dy  \frac{1}{yx^2}  \tan^{-1}\frac{1}{yx^2}  \\
\sim & cu (T/{\it \Gamma})^{2/3}  x_0^{-4} + ... \\
\end{aligned}
\end{equation}
where we define the lower and upper bound for integrals as
$x_0= \sqrt{\delta/(T/{\it \Gamma})^{1/2}}, \ x_c=q_c/(T/{\it \Gamma})^{1/3}$.
In low temperature limit ${T\rightarrow 0}$, 
the upper bound ${x_{\rm c}\sim q_{\rm c}/T^{\frac{1}{2}}}$ 
extends to a large value despite $q_{\rm c}$ is relatively small;
The lower bound $x_0(T)$ also diverges in at low temperatures
given that $\alpha< 1$ (see Eq.~\ref{delta_alpha}).
During the calculation, we retain only the leading order term in the low-temperature limit
which yields the relation 
\begin{equation}
\delta_1(T)\sim  T^{2/3} \frac{T}{\delta^2(T)}.
\label{SM_delta1}
\end{equation}

The second term $\delta_2$ in Eq.~\eqref{SM_delta_delta_0} has an implicit
temperature dependence through $\delta(T)$.
This fact can be seen from,
\begin{equation}
\begin{aligned}
 \delta_2= &  \frac{cu}{2\pi} \int \frac{d^3{\bm  q}}{(2\pi)^3} {\it \Gamma}_{q} \int_0^{1} ds \Big\{ \frac{s}{(\delta+r_{\rm c}-Dq+q^2)^2+s^2}  \\
&-  \frac{s}{(r_{\rm c}-Dq+q^2)^2+s^2}  \Big\}  \\
\sim & u \int d^2{\bm  q} {\it \Gamma}_{q} \ln\Big|\frac{q_0^2-Dq+q^2}{\delta+q_0^2-Dq+q^2} \Big| \\
= & u{\it \Gamma}  \int_0^{q_{\rm c}} d(\delta q_{\parallel}) \int_0^{q_{\rm c}} d(\delta q_\perp)\sqrt{(\delta q_{\parallel})^2 + (\delta q_{\perp})^2}\ln\frac{(\delta q_{\parallel})^2}{\delta+(\delta q_{\parallel})^2} \\
= & u{\it \Gamma}  \int_0^{q_{\rm c}} d(\delta q_{\parallel}) \
\Big\{ \frac{q_c}{2}\sqrt{q_c^2+(\delta q_{\parallel})^2} \\
& + \frac{1}{2}(\delta q_{\parallel})^2\ln\frac{q_c+\sqrt{q_c^2+(\delta q_{\parallel})^2}}{\delta q_{\parallel}} \Big\}\ln\frac{(\delta q_{\parallel})^2}{\delta+(\delta q_{\parallel})^2}.\\
\end{aligned}
\end{equation}
To extract the $\delta$ dependence,
we perform a transformation ${x=\delta q/\delta^{\frac{1}{2}}}$.
The integrals are truncated at a upper bound ${x_{\rm c}\sim {\delta}^{-1/2}}$,
and the integrals are evaluated approximately at large $x$,
\begin{equation}
\begin{aligned}
\delta_2 \sim & u \delta \int_0^{x_{\rm c}} dx \frac{\sqrt{x_c^2+x^2}}{x^{2}} \\
&+u \delta^{3/2} \int_0^{x_{\rm c}} dx \  \ln\frac{x_c+\sqrt{x_c^2+x^2}}{x}  \\
\sim &\ u\delta +  u\delta^{3/2} x_c  \sim u \delta . 
\end{aligned}
\label{SM_delta2}
\end{equation}

Combining the results from Eq.~\eqref{SM_delta1} and Eq.~\eqref{SM_delta2},
we deduce that
\begin{equation}
\delta(T) = T^{\alpha}, \ \alpha=\frac{5}{9} .
\label{SM_delta_T}
\end{equation}\\

The above result with $\alpha<1$ justifies our assumption.
Moreover, let's try to find a solution for $\alpha>1$.
An important difference is that $\delta/T \rightarrow 0$ in the low-temperature limit
so that the right-hand side of Eq.(\ref{SM_delta1}) has no explicit dependence on $\delta(T)$.
This fact simplifies the calculation and yields
\begin{equation}
\begin{aligned}
\delta_1
\simeq & \frac{cu}{2}  \int \frac{d^2{\bm  q}}{(2\pi)^2} \frac{2T}{{\it \Gamma}_{q}}  \int_0^{T}   \frac{{d\epsilon}/{\pi}}{(\delta+r_{\rm c}-D q+q^2)^2+(\epsilon/{\it \Gamma}_{q})^2}   \\
= &  \frac{cu}{2}  \int \frac{d^2{\bm  q}}{(2\pi)^2} \frac{2T}{{\it \Gamma}_{q}} \frac{1}{|\delta+(q-q_0)^2|} \tan^{-1}\frac{T/{\it \Gamma}_{q}}{|\delta+(q-q_0)^2|}   \\
\simeq & cu \int \frac{d(\delta {\bm q}_\parallel)}{(2\pi)}\int \frac{d(\delta {\bm q}_\perp)}{(2\pi)}  \frac{T/{\it \Gamma}}{\sqrt{(\delta q_\parallel)^2+(\delta q_\perp)^2}} \\
& \times \frac{1}{(\delta q_\parallel)^2} \tan^{-1}\frac{(T/{\it \Gamma}) }{\sqrt{(\delta q_\parallel)^2+(\delta q_\perp)^2} (\delta q_\parallel)^2}   \\
= & \frac{cu}{4\pi^2}(T/{\it \Gamma})^{2/3}  \int^{x_c} dx dy  \frac{1}{x^2\sqrt{x^2+y^2}} \tan^{-1}\frac{1 }{x^2\sqrt{x^2+y^2}}   \\
\sim & u T^{2/3} \\
\end{aligned}
\end{equation}
here the cutoff for $x,y$-integrals can be taken to infinity $x_c=q_c/(T/{\it \Gamma})^{1/3} \xrightarrow{T\rightarrow 0} +\infty$.
And, the obtained exponent $\alpha =2/3$ is inconsistent with the assumption $\alpha>1$
so that we conclude that there's no solution for the saddle point equation with $\delta(T) = T^{\alpha}, \ \alpha> 1$.

\section{Scaling analysis}

In this appendix, we make use of the renormalized boson polarization 
for the lowest critical mode
and conduct a scaling analysis for all the parameters 
and field operators in the low-energy effective theory.
We demonstrate the existence of stable Gaussian fixed point
where the fermions and bosons are decoupled,
thereby, the Hertz-Millis method and the corresponding SCR scheme can be safely applied.

We elaborate on the validity of applying
the Hertz-Millis method to the critical boson surface-induced criticality.
We start from the fermion-boson coupled model in Eq.(\ref{eq1}) 
and conduct a renormalization group study.
On the tree level, we analyze the scaling dimensions
of the parameters and field operators in the coupled model.
We show that the model flows to a weakly coupled Gaussian fixed point, 
where the Yukawa coupling constant $g$ is irrelevant.
The validity of the Hertz-Millis method relies on the boson and fermion sectors
are described by respective Gaussian theories.
The four-fermion interaction is irrelevant in obeying the picture of the usual Fermi liquid theory;
however the quartic boson interaction $u$ is marginal at tree level,
while higher-order interactions all become more irrelevant.
This indicates the system is at the upper critical dimension $d_{\rm c}=3$
where the Hertz model (particularly the $\phi^4$ expansion) 
may be inadequate to describe the QCP due to additional mechanisms.
The boson vertices can be drastically changed by anomalous nonlocal contributions
which have been overlooked in the original Hertz-Millis theory\cite{Abanov2004}.
This classical paper is devoted to the case of a gapless boson at ${\bm q}\ne 0$;
while, for our case with the critical boson surface $|{\bm q}|=q_0$,
we show that these dominant nonlocal vertices are irrelevant in all orders,
and the Hertz-Millis scheme still applies.

\subsection{Gaussian theories}

To be specific, we diagonalize the Gaussian part of the boson theory in Eq.(\ref{phi_4})
and obtain the eigen-energies ${\cal E}_n({\bm q},i\omega_l) = f({\bm q},i\omega_l) + nDq$ 
and eigenvectors ${\vec \phi}_n = \phi_n \hat{e}_n$ with the subindex being $n=-1,0,+1$.
The lowest mode ${\vec \phi}_{-1}$ reaches quantum criticality
and is responsible for the induction of NFL.
We project the fermion-boson coupled model into the lowest bosonic mode
and abbreviate $\phi_{-1}$ as $\phi$.
The effective low-energy boson theory at the QCP is written as
\begin{equation}
\begin{aligned}
S_{\rm B}^{(2)} = & \sum_{{\bm q},\omega_l} \Big[ (|{\bm q}|-q_0)^2 + \frac{|\omega_l|}{{\it \Gamma}_q} \Big]  
\phi({\bm q},i\omega_l) \phi(-{\bm q},-i\omega_l),\\
\end{aligned}
\end{equation}
which can be further expanded with respect to small momentum variations in the vicinity of the CBC
\begin{equation}
\begin{aligned}
& (|{\bm q}|-q_0) \simeq \delta q_\parallel + \frac{\delta q_\perp^2}{2q_0}, \\
& {\it \Gamma}_q \simeq {\it \Gamma} \delta q .\\
\end{aligned}
\label{SCR_eq1}
\end{equation}
Similarly, we retain the coupling between the fermion bilinear to the lowest bosonic mode 
in the Yukawa interaction and omit the subindex.
The effective Yukawa interaction reads,
\begin{equation}
\begin{aligned}
S_{\rm BF} = &g \sum_{{\bm q},\omega_l}\sum_{{\bm k},\epsilon_n} \psi^{\dagger}_\alpha({\bm k}+{\bm q},i\epsilon_n + i\omega_l) {\hat \sigma}_{\alpha\beta} \psi_\beta({\bm k},i\epsilon_n) \cdot {\vec \phi}({\bm q},i\omega_l). \\
\end{aligned}
\label{SCR_BF}
\end{equation}

We note that the 2nd order term in Eq.(\ref{SCR_eq1}) 
is a manifestation of the boson surface curvature
at a given momentum ${\bm q}_0$. 
A similar curvature effect is manifested in Fermi liquid theory around the Fermi surface;
yet, we treat the Fermi surface in a flat limit compared to the bosonic one
due to the fact that $q_0\ll k_F$.
This limit is consistent with the calculations conducted in previous sections.
As a result, the curvature term from the Gaussian boson theory
determines the scaling dimension of the spacetime,
\begin{equation}
\begin{aligned}
& x_{\parallel}^\prime =x_{\parallel} e^{-l}, \\
& x_{\perp}^\prime =x_{\perp} e^{-\gamma l}, \\
& \tau^\prime = \tau e^{-zl}. \\
\end{aligned}
\end{equation}
Here we use a generic parameter $1/2\le \gamma \le 1$ 
for the scaling dimension of the perpendicular components.
Two terms in Eq.(\ref{SCR_eq1}) scale in the same way for $\gamma=1/2$.
The Gaussian theory is invariant under spacetime rescaling,
giving rise to the scaling dimension for the boson field
\begin{equation}
{\rm dim}[\phi^2] = -[3+z+(d-1)\gamma]
\label{SCR_eq3}
\end{equation}

The Gaussian part of the Fermi liquid theory can be treated accordingly,
\begin{equation}
\begin{aligned}
{\cal S}_{\rm F}^{(2)} = & \int d\tau d^{d}{\bm x}
\psi^\dagger \Big\{\zeta \partial_\tau - i v_F \hat{k}_F\cdot\partial_{{\bm x}}  \Big\} \psi \\
\end{aligned}
\end{equation}
where the quadratic dispersion is linearized by preserving only the 1st-order terms;
the 2nd order terms, conceiving the curvature of Fermi surface, are ignored.
The linear term is projected on the the $\hat{k}_F$-direction,
which is perpendicular to the boson vector ${\bm q}_0$ as explained around Fig.~\ref{Fig5}.
Therefore, we adapt the consistent notations for coordinates 
and momentums [see {\sl e.g.} Eq.(\ref{q_decomp})],
and we further decompose the tangential components as
\begin{equation}
\begin{aligned}
{\bm x} = & (x_\parallel,  x_\perp), \ \
x_\perp =\hat{k}_F \cdot {\bm x} , \ \  x_\parallel = \hat{q}_0 \cdot {\bm x}. \\
\end{aligned}
\label{SCR_eq6}
\end{equation}

Tree-level RG analysis tells us that 
the parameter $\zeta$ introduced by hand has a scaling dimension
${\rm dim}[\zeta] = -(z-\gamma)<0$ which can be neglected henceforth.
And, the fermion field operator acquires a scaling dimension
\begin{equation}
{\rm dim}\big[\psi({\bm x},\tau)\big] = [z+1+(d-2)\gamma]/2
\label{SCR_eq4}
\end{equation}
from this, we can deduce that the four-fermion interaction is irrelevant,
\begin{equation}
\begin{aligned}
{\cal S}_{\rm F}^{(4)} = & u_0  \int d\tau d^{d}{\bm x}\  \Big\{ \psi^\dagger \psi^\dagger \psi \psi \Big\} \\
{\rm dim}[u_0] = & -(z+1+d)+3\gamma <0 \\
\end{aligned}
\end{equation}
which justifies the usual Fermi liquid picture.

\subsection{Boson quartic interaction}

The boson quartic interaction for the lowest mode takes a form,
\begin{equation}
{\cal S}^{(4)}_{\rm B} = \frac{u^{4}}{4} \int d\tau d^{d}{\bm x} \big[\phi({\bm x},\tau)\big]^4 
\end{equation}
naive scaling analysis based on this expression gives rise to 
\begin{equation}
{\rm dim}[u^{4}]= 3-z-(d-1)\gamma 
\end{equation}
which is irrelevant for our case $z=3, d=2$
regardless of the value of $\gamma$.
Moreover, all higher order boson interactions are irrelevant, {\sl e.g.} 
${\rm dim}[u^{6}]= 4-2[z+(d-1)\gamma] <0$.
It is convinced that our system in the main text is above the upper critical dimension
and the Ginzburg-Landau expansion is faithful.

\subsection{Fermion-boson coupling}

The previous analysis focus on respective Gaussian fixed points of the Hertz theory,
which relies on the irrelevance of the boson and fermion interactions.
The Yukawa coupling between bosons and fermions is equally important in the RG analysis.
Recalling the scaling dimensions of the fermion [Eq.(\ref{SCR_eq4})] and boson [Eq.(\ref{SCR_eq3})] fields,
we can readily deduce the scaling dimesnion for the Yukawa coupling constant
\begin{equation}
{\rm dim}[g] = -\frac{z-1+(d-3)\gamma}{2} .
\end{equation}
The Yukawa coupling is irrelevant ${\rm dim}[g] <0$
for our case $d=2,\ z=3$ regardless of the value of $\gamma$.
Similar to the boson vertex [see Eq.(\ref{SCR_eq5})], 
we notice that the curvature parameter $\gamma$ 
doesn't play a role in the cubic vertex renormalization at the tree level.
The Gaussian fixed point is stable against the Yukawa coupling between fermions and bosons,
thereby, justifies the validity of the Hertz-Millis method and the corresponding SCR scheme.

\begin{figure}[htbp]
\includegraphics[width=8cm]{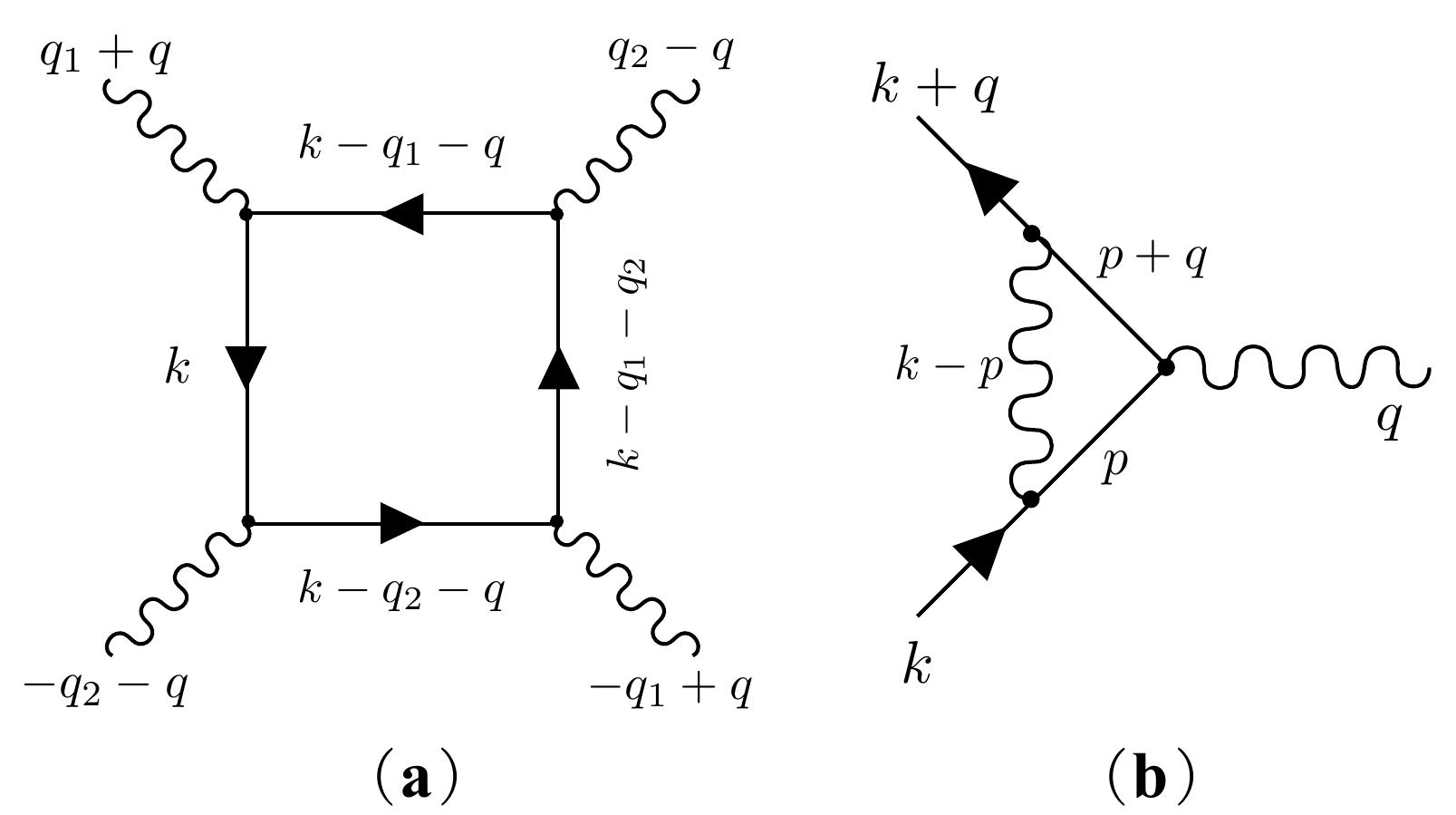}
\caption{ Feynman diagrams for (a) quartic boson vertex; (b) cubic boson-fermion vertex.}
\label{Fig6}
\end{figure}

\subsection{Non-local boson interactions}

There are several additional mechanisms that can fail the Hertz-Millis\cite{RMP2007};
importantly, the local analytic expansion of the bosonic action 
in terms of the magnetic order parameter may break down under certain conditions,
thus, starting point of Hertz-Millis becomes invalid.
A fundamental mechanism is raised by \emph{Abanov and Sachdev}\cite{Abanov2004}.
They show that the Hertz-Millis method of quantum criticality 
is incomplete as it misses anomalous nonlocal contributions to the boson interaction vertices.
For the physically important cases at the upper critical dimension,
the quartic interaction is not the only vertex that becomes marginal.
In fact, the number of marginal vertices is infinite which leads to a dramatic
change in the susceptibility;
infinite many logarithmic corrections add up to a pow-law behavior 
which completely alters the low-energy physics.

In expect of this situation, we follow the procedure in Ref.[\onlinecite{Abanov2004}] and 
calculate anomalous nonlocal contributions to the interaction vertices at all orders $n\ge 2$.
We find that the interaction vertices are irrelevant 
with the presence of the CBC in $d=2$,
regardless of the scaling dimension $\gamma$ brought by the curvature effect. 

To be specific, we calculate the quartic vertex function 
from the Feynman diagram in Fig.~\ref{Fig6}(a).
The simplest case is examined by setting the external boson momentums as
$({\bm q}_1,i\omega_l^1)=({\bm q},i\omega_l)=0,\ ({\bm q}_2, i\omega_l^2)= (-{\bm q}_0,0).$
The $n=2$ vertex correction vanishes due to the presence of 2nd-order poles,
\begin{equation}
\begin{aligned}
b_{4}\sim & \int d\epsilon_n d^3{\bm k}\ \Big\{ G({\bm k},i\epsilon_n) G({\bm k}+{\bm q}_0,i\epsilon_n)  \\
& \times G({\bm k}+{\bm q}_0,i\epsilon_n) G({\bm k},i\epsilon_n) \Big\}, \\
= & \int d\epsilon_n d^3{\bm k} \frac{1}{(i\epsilon_n -\epsilon_{\bm k})^2} 
\frac{1}{(i\epsilon_n -\epsilon_{{\bm k}+{\bm q}_0})^2} =0. \\
\end{aligned}
\end{equation}
To have a finite result, we have to take $({\bm q},i\omega_l)\ne 0$ into account,
which immediately yields 
\begin{equation}
\begin{aligned}
b_4({\bm q},i\omega_l)\sim & \frac{|\omega_l| }{\big[i\omega_l - v_F ({\bm q}-{\bm q}_0)\cdot \hat{k}_F\big]^2} \\
\end{aligned}
\end{equation}
And, the higher-order interaction vertices can be evaluated in a similar fashion by 
preserving the external momentum and frequency $({\bm q},i\omega_l)$ 
[see Fig.2(b)\cite{Abanov2004} for diagrams of higher order boson vertices].
This rough estimation is enough for the purpose of scaling dimension counting.
The interaction at all orders is expressed symbolically by
\begin{equation}
\begin{aligned}
\Gamma^{2n}_{\rm B} = & u^{2n} \int (d^3{\bm q} d\omega_l)^{2n-1} b_{2n}({\bm q},i\omega_l) \big[ \phi({\bm q},i\omega_l)^2 \big]^n  \\
b_{2n}({\bm q},i\omega_l)\sim & \frac{|\omega_l| }{\big[i\omega_l - v_F ({\bm q}-{\bm q}_0)\cdot \hat{k}_F\big]^{2(n-1)}} \\
\end{aligned}
\end{equation}
As argued by \emph{Abanov and Sachdev},
these new terms dominate over the regular $\phi^4$ expansions.
From the Gaussian theory for bosons, we have already known that
$ {\rm dim}[\delta {\bm q}_\perp] =\gamma,\  {\rm dim}[\omega_l] = z,\ 
{\rm dim} \big[ \phi^2(\delta {\bm q},i\omega_l) \big] = -[3+z+(d-1)\gamma]$.
The coefficient of $n$'th order interaction, therefore, has a scaling dimension
\begin{equation}
{\rm dim}[u^{2n}] = - \big[ z-1 +(d-3)\gamma \big]n + \big[1+(d-3)\gamma \big]
\label{SCR_eq5}
\end{equation}
For the case $d=2, z=3$, the coefficients of the non-local boson interaction 
are irrelevant in all orders
and the Ginzburg-Landau expansion remains a legitimate starting point for the Hertz-Millis theory.

\section{Evaluation of frequency dependence of fermionic self-energy}

\begin{widetext}
The integrals in Eq.(\ref{Sig_T0}) is carried out explicitly by using the polar coordinate
\begin{equation}
\begin{aligned}
\Sigma({\bm k}_F,i\epsilon_n;T=0)  
= & g^2  \int\frac{d\omega_l}{2\pi}  \frac{d(\delta {\bm q}_{\parallel})}{2\pi} \frac{d(\delta {\bm q}_\perp)}{2\pi}
\Big\{ (\delta q_\parallel)^2+ \frac{|\omega_l|}{\Gamma \delta q}  \Big\}^{-1} 
\frac{1}{i(\epsilon_n+\omega_l)- v_F \delta q_\perp} \\
= & -i\frac{g^2 v_F}{4\pi^2} \int d\omega_l \int_0^{+\infty} dq \frac{1}{q^2} \int_0^{2\pi} \frac{d\phi}{2\pi} 
\frac{1}{\sin^2 \phi + \frac{|\omega_l|}{\Gamma q^3}}  \frac{(\epsilon_n+\omega_l)/(v_Fq)}{[(\epsilon_n+\omega_l)/(v_F q)]^2 + \cos^2\phi} \\
\simeq & -i\frac{g^2 }{4\pi^2 v_F} \int d\omega_l \int_0^{+\infty}  \frac{dq}{q^2} \Big\{ \frac{\epsilon_n+\omega_l}{v_Fq}  f\big(|\omega_l|/\Gamma q^3 \big) 
+ \frac{1}{2}\frac{\big[{\rm sgn}(\epsilon_n+\omega_l)+1\big]}{1+ |\omega_l|/\Gamma q^3} \Big\}.\\
\end{aligned}
\label{SM_Sig_T0}
\end{equation}
We have used the integral with $a= \sqrt{|\omega_l|/(\Gamma q^3)},\ b=(\epsilon_n+\omega_l)/(v_F q) $.
The result is approximated in the limits $a\ll b\ll 1$, which yields 
\begin{equation}
\begin{aligned} 
\int_0^{2\pi} \frac{d\phi}{2\pi} \frac{1}{\sin^2\phi + a^2} \frac{1}{\cos^2\phi+ b^2}
 = & \frac{1}{a\sqrt{1+a^2}(1+a^2+b^2)} + \frac{\big[{\rm sgn}(b)+1\big]/2}{b\sqrt{1+b^2}(1+a^2+b^2)} \\
\simeq &f(a) + \frac{\big[{\rm sgn}(b)+1\big]/2}{b(1+a^2)} .\\
\end{aligned}
\end{equation}
where the function $f(a) =  \frac{1}{a\sqrt{1+a^2}(1+a^2)}$.\\

The two terms in Eq.(\ref{SM_Sig_T0}) are calculated separately.
The 1st term is calculated by making a change of variable $x=\omega_l/\Gamma q^3,\ y =q/(\epsilon_n/\Gamma)^{1/3}$,
which reads
\begin{equation}
\begin{aligned}
 \sim &   \int d\omega_l \int_0^{+\infty} dq \frac{1}{q^2} \frac{\epsilon_n+\omega_l}{v_Fq}  f\big(|\omega_l|/\Gamma q^3 \big)
 = \frac{\Gamma}{v_F} \int_0^{+\infty} dq  \Gamma q^3 \int dx \Big\{ \frac{\epsilon_n}{\Gamma q^3 } + x\Big\}  f\big(|x| \big) \\
& = \frac{\Gamma^2}{v_F} (\epsilon_n/\Gamma)^{4/3} \int_0^{+\infty} dy  \int dx \Big\{ \frac{1}{y^3} + x\Big\}  f\big(|x| \big)
\equiv   \frac{c_1 \Gamma^{2/3}}{v_F} \epsilon_n^{4/3},  \\
\end{aligned}
\label{SM_Sig_T0_1}
\end{equation}
where the constant is given by $c_1= \int_0^{+\infty} dy \int dx (y^{-3} + x)  f(|x|) $.
The 2nd term is calculated by making a change of variable $y =q/(|\epsilon_n|/\Gamma)^{1/3}$
\begin{equation}
\begin{aligned}
  \sim  &\frac{1}{2} \int d\omega_l \int_0^{+\infty} dq \frac{1}{q^2}  \frac{{\rm sgn}(\epsilon_n+\omega_l)}{1+ |\omega_l|/\Gamma q^3}
  =  \frac{1}{2} \int d\omega_l \int_0^{+\infty} dq   \frac{{\rm sgn}(\epsilon_n+\omega_l)}{q^2+ |\omega_l|/\Gamma q} \\
&=  2{\rm sgn}(\epsilon_n) \int_0^{+\infty} dq \Gamma q \ln \frac{q^2+ |\epsilon_n|/\Gamma q}{q^2} 
= 2\Gamma (|\epsilon|/\Gamma)^{2/3} {\rm sgn}(\epsilon_n) \int_0^{+\infty} dy y \ln \frac{y^3+1}{y^3}  \\
& =  c_2  \Gamma^{1/3}  {\rm sgn}(\epsilon_n) |\epsilon_n|^{2/3},\\
\end{aligned}
\label{SM_Sig_T0_2}
\end{equation}
where the constant is given by $c_2=\int_0^{+\infty} dy y \ln (1+ y^{3})/y^3$.\\

We note that the 2nd term in Eq.(\ref{SM_Sig_T0_2}) dominates over the 1st term in Eq.(\ref{SM_Sig_T0_1})
in the low-energy limit.
Finally, we arrive at the conclusion that the leading contribution to the self-energy correction at zero temperature is given by
\begin{equation}
\begin{aligned}
\Sigma({\bm k}_F,i\epsilon_n;T=0)  \simeq -ic_2\frac{g^2  \Gamma^{1/3}}{4\pi^2 v_F}{\rm sgn}(\epsilon_n) |\epsilon_n|^{2/3}.
\end{aligned}
\label{SM_Sig_T0_3}
\end{equation}

\end{widetext}

\noindent{{\bf Acknowledgments}}\\
Not applicable.
\\

\noindent{{\bf Funding}}\\
This work is supported by the National Science Foundation of China with Grant No. 92065203, the Ministry of Science and Technology of China with Grants No. 2018YFE0103200, by the Shanghai Municipal Science and Technology Major Project with Grant No. 2019SHZDZX04, and by the Research Grants 
Council of Hong Kong with General Research Fund Grant No. 17306520.\\

\noindent{{\bf Availability of data and materials}}\\
All data and materials are included in the main text and appendices.
Further information can be requested from corresponding author via e-mail.\\

{\large \noindent{{\bf Declarations}}}\\

\noindent{{\bf Competing interests}\\
The authors declare no competing interests.\\

\noindent{{\bf Author contributions}\\
G.C. designed and supervised this project. 
X.T.Z. and G.C. performed the calculation and wrote the manuscript.

\bibliography{Ref.bib}

\end{document}